\documentstyle[12pt,epsf,floats,preprint,prd,aps]{revtex}

\tighten
\newlength{\notewidth}
\newcommand{\kkm}{${\text{ K}^0\!-\!\overline{\text{K}^0}}\,$-mixing\/}
\newcommand{\bbm}{${\text{ B}^0\!-\!\overline{\text{B}^0}}\,$-mixing\/}
\newcommand{\bbmd}{${\text{B}_d^0\!-\!\overline{\text{B}^0_d}}\,$-mixing\/}
\newcommand{\bbms}{${\text{B}_s^0\!-\!\overline{\text{B}^0_s}}\,$-mixing\/}

\newcommand{\kkmd}{${\text{K}_{\text{L}}\!-\!\text{K}_{\text{S}}}\,$-mass
difference\/}
\newcommand{\dst}{|\Delta \text{S}|\text{=2}}
\newcommand{\hdst}{H^{|\Delta \text{S}| =  2}}
\newcommand{\dso}{|\Delta \text{S}|\!=\! 1}
\newcommand{\hdso}{H^{|\Delta \text{S}|  =  1}}
\newcommand{\laMSb}{\Lambda_{\overline{\text{MS}}}^{\text{NLO}}}
\newcommand{\rf}[1]{(\ref{#1})}
\newcommand{\ov}{\overline}
\newcommand{\np}{Nucl.\ Phys.\ }
\newcommand{\pr}{Phys.\ Rev.\ }
\newcommand{\prp}{preprint }

\newcommand{\gev}{\, \text{GeV}}
\newcommand{\imag}{\text{Im}\,}
\newcommand{\real}{\text{Re}\,}
\newcommand{\ek}{\varepsilon_K}

\begin{document}
\draft
\preprint{\parbox[t]{4.5cm}
          {TUM-T31-81/95 \\
           PSI-PR-95-13 \\
           hep-ph/9507262 \\
           July 1995}}
\title{\Large
	Indirect CP-Violation in the Neutral Kaon System Beyond
	Leading Logarithms
	\thanks{Work supported by BMBF under contract no.\ 06-TM-743.}
      }
\author{Stefan Herrlich
        \footnote{\vspace{-2mm}e-mail:
         {\tt herrl@feynman.t30.physik.tu-muenchen.de}}}
\address{\sl Paul Scherrer Institut, CH-5232 Villigen PSI, Switzerland}
\author{Ulrich Nierste
        \footnote{\vspace{-2mm}e-mail:
	 {\tt nierste@feynman.t30.physik.tu-muenchen.de}}}
\address{\sl Physik-Department, TU M\"unchen, D-85747 Garching, Germany}
\maketitle
\begin{abstract}
We have calculated the short distance QCD coefficient $\eta_3$ of the
effective $|\Delta \text{S}|\text{=2}$-hamiltonian in the
next-to-leading order of renormalization group improved perturbation
theory.  Since now all coefficients $\eta_1$, $\eta_2$ and $\eta_3$
are known beyond the leading log approximation, one can achieve a much
higher precision in the theoretical analysis of $\varepsilon_K$, the
parameter of indirect CP-violation in \kkm.  The measured value for
$\varepsilon_K$ yields a lower bound on each of $|V_{cb}|$,
$|V_{ub}/V_{cb}|$, the top quark mass $m_t$ and the non-perturbative
parameter $B_K$ as a function of the remaining three quantities.
E.g.\ $m_t^{\text{pole}}=176\gev$, $|V_{cb}|=0.040 $ and $B_K=0.75$
implies $|V_{ub}/V_{cb}| \geq 0.0778$, if the measured value for
$\varepsilon_K$ is attributed solely to Standard Model physics.  We
further discuss the implications on the CKM phase $\delta$,
$\left|V_{td}\right|$ and the key quantity for all CP-violating
processes, $\imag \lambda_t = \imag \left[ V^*_{ts} V_{td} \right]$.
These quantities and the improved Wolfenstein parameters $\bar\rho$
and $\bar\eta$ are tabulated and the shape of the unitarity triangle
is discussed.  We compare the range for $|V_{td}|$ with the one
obtained from the analysis of \bbmd.  For $0.037 \leq
\left|V_{cb}\right| \leq 0.043$, $0.06 \leq \left|V_{ub}/V_{cb}\right|
\leq 0.10 $ and $0.65 \leq B_K \leq 0.85$ we find
from a combined analysis of $\ek$ and the \bbmd\/ paramater
$x_d$:
$49^{\circ} \leq \delta \leq 146^{\circ}$,
$7.4 \cdot 10^{-3} \leq \left|V_{td}\right| \leq 12.4 \cdot 10^{-3} $,
$ 0.85 \cdot 10^{-4} \leq \imag \lambda_t \leq 1.60 \cdot 10^{-4}$,
$-0.36 \leq \bar\rho \leq 0.28$ and $0.21 \leq \bar\eta \leq 0.44$.
We predict the mass difference of the $B_s^0$ system to lie in the
range $6.5 \, ps^{-1} \leq \Delta m_{B_s} \leq 28 \, ps^{-1} $.
Finally we have a 1995 look at the
\kkmd.
\end{abstract}
\setcounter{footnote}{0}
\renewcommand{\thefootnote}{\arabic{footnote})}

\section{Introduction}
Since its discovery in the year 1964 \cite{ccft} the study of
CP-violation is of continuous interest to both experimentalists and
theoreticians.  The Standard Model mechanism of CP-violation involves
only a single parameter, the phase $\delta$ in the
Cabibbo-Kobayashi-Maskawa (CKM) matrix.  Hence first the investigation
of CP-violating processes is a useful tool in the determination of the
CKM elements, some of which are poorly known at present.  Second it
may be the key to find physics beyond the Standard Model, once one
will not be able to fit different observables with the single
parameter $\delta$.

Yet at present CP-violation is only precisely and unambiguously
measured in
$\dst$-tran\-sitions.
It manifests itself in the fact that
the neutral Kaon mass eigenstates $|\text{K}_{\text{L}}\rangle$ and
$|\text{K}_{\text{S}}\rangle$ are no CP eigenstates.  This indirect
CP-violation is characterized by the parameter
\begin{eqnarray}
\varepsilon_K &=& \frac{\langle ( \pi \pi )_{I=0} | \hdso  |
                  \text{K}_{\text{L}} \rangle }
              {\langle ( \pi \pi )_{I=0} | \hdso  | \text{K}_{\text{S}}
              \rangle } .
\end{eqnarray}
Its relation to the low-energy $\dst $-hamiltonian $\hdst$ is given
(in the CKM phase convention for $| \text{K}^0 \rangle $) by
\begin{eqnarray}
\varepsilon_K &=& \frac{e^{i \pi/4}}{\sqrt{2}}
         \left( \frac{\mbox{Im\,} \langle
           \text{K}^0 | \hdst  | \overline{\text{K}^0}
            \rangle }{\Delta m_K} + \xi \right) . \label{ek}
\end{eqnarray}
Here $m_K$ is the neutral Kaon mass, $\Delta m_K$ is the \kkmd\/ and
$\xi $ is a small quantity related to CP violation in the $\dso$
amplitudes, it contributes roughly 3\% to $|\varepsilon_K|$ (see
\cite{cbh} for details).

The theorist's challenge is the proper inclusion of the strong
interaction, which binds the quarks into hadrons and screens or
enhances the CP-violating weak amplitude.  Here the short distance QCD
effects can be reliably calculated in renormalization group (RG)
improved perturbation theory.  With our new calculation they are now
completely known in the next-to-leading order (NLO).  Its
phenomenological implications are the subject of this paper, which is
organized as follows: In the following section we present the
$\dst$-hamiltonian in the NLO.  The further ingredients of the
phenomenological analysis are discussed in sect.~\ref{input}.  In
sect.~\ref{borderline} we analyze which region of the Standard Model
parameters is compatible with the observed value for $\varepsilon_K$.
In sect.~\ref{vtd} we first determine the CKM phase $\delta$ from
$\varepsilon_K$. Then we obtain $|V_{td}|$, which is a key quantity
for \bbmd, and discuss the additional constraints obtained from the
measured \bbmd\/ mixing parameter $x_d$. Further we determine the
improved Wolfenstein parameters $\bar\rho$ and $\bar\eta$ and further
$\imag \lambda_t$, which is proportional to the Jarlskog measure of
CP-violation and therefore enters all CP-violating quantities in the
Standard Model.  Finally we discuss the short distance contributions
to the \kkmd.

\section{The \protect$\dst $-hamiltonian in the next-to-leading order}
The low-energy hamiltonian inducing \kkm\/ reads:
\begin{eqnarray}
H^{|\Delta S|=2} &=&
               \frac{ G_{F}^2 }{ 16 \pi^2 } M_W^2  \left[
                  \lambda_c^2 \eta_1
                   x_c  \! + \!
                  \lambda_t^2 \eta_2
                  S(x_t ) \! + \!
                2 \lambda_c \lambda_t \eta_3
                  S(x_c , x_t  )
                   \right]
 b(\mu) Q_{S2}(\mu) + \text{h.c.} \; \; \label{s2}
\end{eqnarray}
Here $G_{F}$ is the Fermi constant, $M_W$ is the W boson mass and $x_i
=m_i^2 /M_W^2$.
\begin{eqnarray}
\lambda_j=V_{jd} V_{js}^{*} \label{laj}
\end{eqnarray}
comprises the
CKM-factors and $Q_{S2} $ is the local four-quark operator
\begin{eqnarray}
Q_{S2} &=&
( \overline{s}_j \gamma_\mu (1-\gamma_5) d_j)
(\overline{s}_k \gamma^\mu (1-\gamma_5) d_k) \; = \; (\overline{s} d)_{V-A}
(\overline{s} d)_{V-A}  \label{ollintro}
\end{eqnarray}
with $j$ and $k$ being colour indices.  The Inami-Lim functions
\cite{il}
\begin{eqnarray}
S(x_t) &=& x_t \left[ \frac{1}{4} + \frac{9}{4}
\frac{1}{1-x_t} - \frac{3}{2} \frac{1}{(1-x_t)^2}      \right]
- \frac{3}{2} \left[ \frac{x_t}{1-x_t} \right]^3 \ln x_t \nonumber \\
S(x_c,x_t) &=& - x_c \ln x_c
+x_c \left[ \frac{x_t^2-8 x_t+4 }{4 (1-x_t)^2} \ln x_t
          + \frac{3}{4} \frac{x_t}{x_t-1}   \right] 
\end{eqnarray}
depend on the masses of the charm- and top-quark and describe the
$|\Delta \text{S}|\text{=2} $-transition amplitude in the absence of strong
interaction.

The short distance QCD corrections are comprised in the coefficients
$\eta_1$, $\eta_2$ and $\eta_3$ with a common factor $b(\mu)$ split
off.  They are functions of the charm and top quark masses and of the
QCD scale parameter $\Lambda_{\text{QCD}}$.  Further they depend on
various renormalization scales.  This dependence, however, is
artificial, as it originates from the truncation of the perturbation
series, and diminishes order-by-order in $\alpha_s$.  The $\eta_i$'s
have been calculated in the leading-logarithmic approximation by
Gilman and Wise \cite{gw} for the case of a light top quark. The
corresponding results for a heavy top quark have been derived in
\cite{fp}.  We briefly recall the motivation for the calculation in
the NLO:
\begin{enumerate}
\renewcommand{\labelenumi}{\roman{enumi})}
\item To make use of the fundamental QCD scale parameter
$\Lambda_{\overline{\text{MS}}}$ one must calculate beyond the
leading order (LO).
\item The quark mass dependence of the $\eta_i$'s is not
accurately reproduced by the LO expressions.  Especially the
$m_t$-dependent terms in $\eta_3 \cdot S(x_c,x_t)$ belong to the NLO.
\item The LO results for $\eta_1$ and $\eta_3$
show a large dependence on the
renormalization scales, at which one integrates out heavy particles.
In the NLO these uncertainties are reduced.
\item One must go to the NLO to judge whether perturbation theory
works, i.e.\ whether the radiative corrections are small.  After all
the corrections can be sizeable.
\end{enumerate}
In the NLO one has to take care of the proper definition of the quark
masses. It is most useful to define the $\eta_i$'s with respect to
running masses in the $\overline{\text{MS}}$ scheme normalized as
$m_i^{\star}=m_i(m_i)$.
I.e.\ we use $x_i^{\star } = \left[ m_i (m_i) \right] ^2
/M_W^2$ in \rf{s2} and mark the corresponding $\eta_i$'s with a star.
The NLO calculation here requires the use of  the one-loop relation
between the pole mass and the running mass:
\begin{eqnarray}
m^{\text{pole}} &=& m^{\star}
\left( 1+ \frac{\alpha_s (m^{\star} )}{\pi} \frac{4}{3}   \right).
\nonumber
\end{eqnarray}
The top quark running mass $m_t^{\star}$ is smaller than
$m_t^{\text{pole}}$ by $8 \, \text{GeV}$.

$\eta_2^{\star}$ and $\eta_3^{\star}$ depend very weakly on the charm and top
quark mass and on $\laMSb$, so that they can be treated as constants.
In contrast $\eta_1^{\star}$ is a steep function of
$m_c^{\star}$ and $\laMSb$.

Now the NLO values read:
\begin{eqnarray}
\eta_1^{\star} &=& 1.32
\begin{array}{l}
\scriptstyle +0.21 \\[-.7mm]
\scriptstyle -0.23                 
\end{array}\, , \quad \quad
\eta_2^{\star} \; = \; 0.57
\begin{array}{l}
\scriptstyle +0.00 \\[-.7mm]
\scriptstyle -0.01
\end{array}\, , \quad \quad
\eta_3^{\star} \; = \;  0.47
\begin{array}{l}
\scriptstyle +0.03 \\[-.7mm]
\scriptstyle -0.04                 
\end{array}\, , \label{etas}
\end{eqnarray}
where $m_c^{\star}=1.3 \gev$ and $\laMSb=0.310 \gev$ has been used.
The quoted theoretical errors
are estimated in two ways: First the renormalization scales have been
varied and second the calculated $O(\alpha_s)$--corrections have been squared.

The calculation for $\eta_1^{\star}$ has been performed by us
\cite{hn1} and $\eta_2^{\star}$ has been obtained by Buras, Jamin and
Weisz \cite{bjw}. The NLO value for $\eta_3^{\star}$ in
\rf{etas} is new.
We will present details of the calculation in \cite{hn86}.

For comparison we give the old leading-order central values \cite{gw}:
\begin{eqnarray}
\eta_1^{\star \,\text{LO} } &=& 0.80 \, ,
\quad \quad
\eta_2^{\star \,\text{LO} } \; = \;  0.62 \, ,
\quad \quad
\eta_3^{\star \,\text{LO} } \; = \;  0.36 \, .
\label{old}
\end{eqnarray}
The common factor of  the short distance QCD corrections split off
in \rf{s2} equals
\begin{eqnarray}
b(\mu) &=& \left[ \alpha_s (\mu) \right] ^{ - 2/9 }
         \left( 1 + \frac{307}{162} \frac{\alpha_s(\mu) }{4 \pi}
    \right)      \label{bmu}
\end{eqnarray}
in the NLO. Here $\mu$ is the scale at which the perturbative short
distance calculation is matched to the non-perturbative evaluation of
the hadronic matrix element.
The latter must compensate the $\mu$--dependence in \rf{bmu} and is
parametrized by $B_K$ as
\begin{eqnarray}
\langle \overline{\text{K}^0} | Q_{S2} (\mu) |  \text{K}^0 \rangle &=&
\frac{8}{3} f_K^2 m_K^2 B_K / b( \mu ) .
\label{bk}
\end{eqnarray}
Here $m_K$ and $f_K$ are the mass and decay constant of the neutral
Kaon.

\section{Miscellaneous}\label{input}
\subsection{CKM Matrix and Unitarity Triangle}\label{sectckm}
For all numerical analyses we will use the exact standard
parametrization of the CKM matrix \cite{pdg}:
\begin{eqnarray}
V\; =\;
\left(
\begin{array}{ccc}
V_{ud} & V_{us} & V_{ub} \\
V_{cd} & V_{cs} & V_{cb} \\
V_{td} & V_{ts} & V_{tb}
\end{array}
\right) &=&
\left(
\begin{array}{ccc}
c_{12} c_{13} & s_{12} c_{13} & s_{13} e^{-i \delta} \\
 - s_{12} c_{23} -c_{12} s_{23} s_{13}  e^{i \delta}
& c_{12} c_{23} -s_{12} s_{23} s_{13} e^{i \delta}
& s_{23} c_{13} \\
s_{12} s_{23} - c_{12} c_{23} s_{13} e^{i \delta} &
-c_{12} s_{23} -s_{12} c_{23} s_{13} e^{i \delta} & c_{23} c_{13}
\end{array}
\right)\, ,   \label{exact}
\end{eqnarray}
where $c_{ij}=\cos \theta_{ij}$ and $s_{ij}=\sin \theta_{ij}$.

\begin{figure}[htb]
\centerline{\epsfxsize=8cm
\epsffile{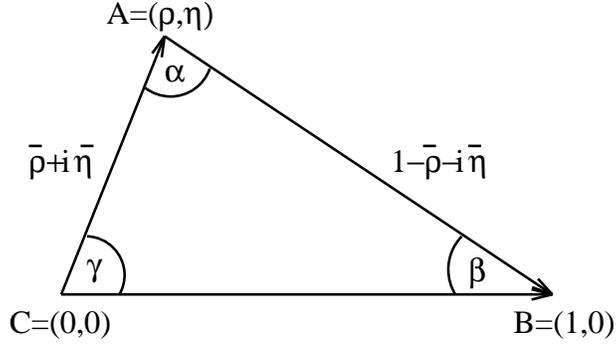}
}
\caption{The unitarity triangle of (\ref{rhobar}).}
\label{unipict}
\end{figure}

The unitarity of $V$ provides us with many relations among its
elements. The most useful one is
\begin{eqnarray}
V_{ud} V_{ub}^* +V_{cd} V_{cb}^* +V_{td} V_{tb}^* &=& 0 \label{uni}.
\end{eqnarray}
With
\begin{eqnarray}
\ov{\rho} &=& - \real \frac{V_{ud} V_{ub}^*}{V_{cd} V_{cb}^*}  ,
\quad \quad \quad
\ov{\eta} \; = \; - \imag \frac{V_{ud} V_{ub}^*}{V_{cd} V_{cb}^*}
\label{rhobar}
\end{eqnarray}
\rf{uni} describes an {\em unitarity triangle}\/
in the complex $\ov{\rho}$--$\ov{\eta}$--plane, whose edges are
located at the points $(0,0)$, $(0,1)$ and $(\ov{\rho},\ov{\eta})$
(see fig.~\ref{unipict}).

To illustrate the size of the contributions from the different CKM
elements we will also use the improved Wolfenstein parametrization
\cite{blo}, which is obtained from \rf{exact}
by defining the parameters $\lambda , A, \rho$ and $\eta$ by
\begin{eqnarray}
s_{12} &=& \lambda \; =\; 0.22, \quad  \quad
s_{23} \; = \; A \lambda^2,  \quad   \quad
s_{13} e^{-i \delta} \; = \;  A \lambda^3 \left( \rho - i \eta \right),
\nonumber
\end{eqnarray}
and expanding the cosines in \rf{exact} to any desired order in
$\lambda=0.22$.  The expansion to order $\lambda^3$ yields the
conventional Wolfenstein parametrization \cite{w}. Yet it is
well-known that the proper treatment of CP-violating effects requires
a higher accuracy:
\begin{eqnarray}
V & =&  \left(
\begin{array}{ccc}
1-\frac{\lambda^2 }{2} &\lambda & A \lambda^3 (\rho - i \eta ) \\
- \lambda -i A^2 \lambda^5 \eta
& 1 -  \frac{\lambda^2 }{2} & A \lambda ^2 \\
A \lambda^3 (1 - \ov{\rho} - i \ov{\eta} )
& -A \lambda^2 -i A \lambda ^4 \eta & 1
\end{array} \right) \label{wolf}
\end{eqnarray}
is exact to order $\lambda^3$ and contains the phenomenologically
important terms up to the order $\lambda^5$ \cite{blo}.  Here
$\ov{\rho}$ and $\ov{\eta}$ defined in \rf{rhobar} are expanded as
\begin{eqnarray}
\ov{\rho } \; = \; \rho \left( 1- \frac{\lambda ^2}{2}
                   +O (\lambda^4) \right) ,
&\quad &
\ov{\eta } \; = \; \eta \left( 1- \frac{\lambda ^2}{2}
                   +O (\lambda^4) \right) .  \nonumber
\end{eqnarray}

\subsection{CKM Elements from $\varepsilon_K$}
The experimental value for $|\varepsilon_K|$ \cite{pdg},
\begin{eqnarray}
| \varepsilon_K | &=& (2.266 \pm 0.023 ) \cdot 10^{-3} ,
\label{expek}
\end{eqnarray}
constrains the CKM elements with \rf{ek}, \rf{s2} and  \rf{bk}
via
\begin{eqnarray}
1.21 \cdot 10^{-7} &=& B_K
\left[ - \imag \lambda_c^2 \, \eta^{\star}_1 \, S(x^{\star}_c)
       - \imag \lambda_t^2 \, \eta^{\star}_2 \, S(x^{\star}_t) -
        2 \, \imag
        \left( \lambda_c \lambda_t \right) \, \eta^{\star}_3 \,
          S(x^{\star}_c,x^{\star}_t) \right] . \label{cons}
\end{eqnarray}
Here the number on the LHS originates from
\begin{eqnarray}
1.21 \cdot 10^{-7}
&=&
 \frac{12 \sqrt{2}\,  \pi^2\, \Delta m_K}{G_F^2 \, F_K^2\, m_K \, M_W^2}
  \left( |\varepsilon_K|   - \frac{\xi}{ \sqrt{2} } \right)
     \nonumber
\end{eqnarray}
with the numerical values for these physical quantities listed in
sect.~\ref{par}.  Further $\lambda_j$ has been defined in \rf{laj} and
the small term $\xi$ in \rf{ek} has been estimated with the help of
\cite{atom} to contribute roughly $-3\%$ to $\varepsilon_K$.  The
uncertainty in the LHS due to experimental errors is about 1\% and
therefore negligible compared to the uncertainties to be discussed in
sect.~\ref{par}.

The relative importance of the three terms in the square bracket in
\rf{cons} can be demonstrated with the help of the improved Wolfenstein
parametrization \rf{wolf} turning \rf{cons} into
\begin{eqnarray}
5.3 \cdot 10^{-4} &=& B_K A^2 \ov{\eta}
    \left[ (1-\ov{\rho}) A^2 \lambda^4 \eta^{\star}_2 S(x^{\star}_t)
           + \eta^{\star}_3 S(x^{\star}_c,x^{\star}_t)
           - \eta^{\star}_1 x^{\star}_c  \right]  \label{cons2}
\end{eqnarray}
after dividing both sides by $2\, \lambda^6$.

In \rf{cons2} one sees that the top-top contribution is CKM suppressed by
four powers of $\lambda$, but this suppression is over-compensated
because the top quark is so heavy:
\begin{eqnarray}
\eta^{\star}_2 S(x^{\star}_t)\; & \approx & \; 1.3 \cdot 10^{3}
\cdot \eta^{\star}_3 S(x^{\star}_c, x^{\star}_t)
\;\; \approx \;\; 4 \cdot 10^3 \cdot  \eta^{\star}_1 x^{\star}_c  . \nonumber
\end{eqnarray}
Hence $\eta^{\star}_2$ is the most important short distance
coefficient, $\eta^{\star}_3$ is second relevant and $\eta^{\star}_1$
contributes least. Their contributions to the RHS of \rf{cons2} are
roughly $75 \%$, $37 \%$ and $-12\%$. Yet if we look at the changes in
the $\eta^{\star}_i$'s due to the NLO calculations (cf.\ \rf{etas} and
\rf{old}) one realizes that the NLO correction to $\eta^{\star}_3$ is
the most important one, because it is enhanced by $30\%$, while
$\eta^{\star}_2$ has decreased by only $8\%$.

\subsection{Ranges for the Input Parameters}\label{par}
In this section we will discuss the actual ranges of the input
parameters needed for our analysis. To determine $\delta$ from
\rf{cons} one must first fix the three angles in
\rf{exact} from the magnitudes of three CKM elements.  While
$|V_{us}|=0.2205 \pm 0.0018 $ is well-known \cite{pdg}, the
determination of $V_{cb}$ and especially $V_{ub}$ from tree-level
b-decays is still plagued by sizeable experimental and theoretical
uncertainties. Since these parameters are two main contributors to our
final error bars, we will now consider them in more detail:

The theoretical understanding of the determination of $V_{cb}$ from
exclusive and inclusive B-decays has recently made significant
progress \cite{bbb}. In \cite{bbb} presumably large perturbative
corrections proportional to $\alpha_s^{n+1} \beta_0^n$ have been
summed to all orders in the decay rate resolving both the previous
discrepancy between the results of inclusive and exclusive analyses
and the large scheme dependence of the inclusive analysis found in
\cite{bn}.  With $\tau_{B_d^0}=(1.59 \pm 0.07) ps$ \cite{p} the
result of \cite{bbb} reads
\begin{eqnarray}
V_{cb} &=& 0.040 \pm 0.003  \label{vcb}
\end{eqnarray}
coinciding with the result presented in \cite{bigi}.  $b \rightarrow
u$ decays are harder to treat both theoretically and
experimentally. We will use \cite{pdg}
\begin{eqnarray}
\left| \frac{V_{ub}}{V_{cb}} \right| &=& 0.08 \pm 0.02 . \label{vub}
\end{eqnarray}

A further ingredient of our analysis is the top quark mass, which has
been determined in the CDF experiment \cite{cdf} to equal
\begin{eqnarray}
m_t^{\text{pole}} &=& (176 \pm 13) \,\mbox{GeV}  . \nonumber
\end{eqnarray}
In NLO analyses one has to take into account the proper definition of
the mass: The corresponding value for the running mass in the
$\ov{\text{MS}}$-scheme is
\begin{eqnarray}
m_t^{\star} \; = \; m_t(m_t) &=& (168 \pm 13) \,\mbox{GeV}  . \label{mass}
\end{eqnarray}
The fit of the top mass from the LEP data yields the same central
value with an error bar of roughly the double size \cite{srie}.  The
D0 group finds $m_t^{\text{pole}}=(199 \pm 30)\, \text{GeV}$
\cite{d0}.  Yet the analysis in \cite{sop} extracting the top mass by
partly fitting the cross sections finds a lower value
$m_t^{\text{pole}}=(170 \pm 9)\, \text{GeV}$ from the combined
analysis of CDF and D0.  Therefore the range given in \rf{mass} well
represents the possible values for $m_t^{\star}$ and will be used in
the following sections.

Next we have to discuss the non-perturbative parameter $B_K$ defined
in \rf{bk}: The size of $B_K$ has been the subject of a controversal
discussion during the last decade. The $1/N_c$ result $B_K=0.7 \pm
0.1$ \cite{bbg} was in contradiction with lower values estimated with
chiral symmetry \cite{gdh} or the QCD hadron duality approach
\cite{pr}.  Yet a recent analysis \cite{bp} has vindicated the result
of \cite{bbg} and seems to have explained the difference to the
estimates in \cite{gdh,pr}.  Further recent quenched QCD lattice
calculations have yielded values around $B_K=0.78$ (see \cite{latt}
and references therein).  The effect of dynamical fermions has been
found to be small in \cite{i}.  We will therefore use the following
range in our calculation:
\begin{eqnarray}
B_K &=& 0.75 \pm 0.10        .  \label{bkrange}
\end{eqnarray}
In fact we will see in sect.~\ref{borderline} that the inclusion of
values lower than $B_K=0.65$ can only very hardly be brought into
agreement with the measured value of $\ek$.  We remark that the NLO
short distance calculation also affects $B_K$ because of the factor of
$1/b(\mu)$ on the RHS of \rf{bk}.  Non-perturbative calculations
determine the matrix element on the LHS of \rf{bk} and usually the
quoted results for $B_K$ are obtained with the leading order factor
$b^{\text{LO}} (\mu) = \left[ \alpha^{\text{LO}} (\mu) \right]
^{-2/9}$ instead of the NLO value given in \rf{bmu}.  Hence in a
consistent NLO analysis one should correct for this by multiplying the
cited values with $b(\mu)/b^{\text{LO}} (\mu)$.  Yet numerically this
amounts to a change of about $3\%$ for $\mu = O \left( 0.7
\,\text{GeV} \right) $ and can be neglected in view of the larger
uncertainty in \rf{bkrange}.  But once the lattice results will
achieve an accuracy in the percentage region they should be quoted
with the NLO factor given in \rf{bmu}.

At this point it is instructive to investigate the impact of our NLO
calculation for $\eta_3^{\star}$: With \rf{cons} one can easily verify
that the shift from $\eta_3^{\star\, \text{LO}}=0.36$ in \rf{old} to
$\eta_3^{\star}=0.47$ in \rf{etas} has the same influence on $|\ek|$
as a shift from $B_K=0.82$ to $B_K=0.75$.  In the same way one can
estimate the uncertainty caused by the error bar in the NLO values in
\rf{etas}: The remaining uncertainties in the NLO $\eta_i^{\star}$'s
correspond to a change in $B_K$ by $\pm 0.02$.

Let us now look at the other input parameters: The dominant QCD
factors $\eta_2^{\star}$ and $\eta_3^{\star}$ depend very weakly on
the QCD scale parameter $\laMSb$, which therefore hardly affects our
results for $\ek$.  Yet of course the determination of the input
parameters $V_{cb}$ and $|V_{ub}/V_{cb}|$ depends on $\laMSb$;
this uncertainty is included in the error bar in
\rf{vcb} and \rf{vub} \cite{bbb}.
Conversely the \kkmd\/ discussed in sect.~\ref{secmass} is dominated
by $\eta_1^{\star}$ which is a steep function of $\laMSb$.  We will
consider \cite{beth}
\begin{eqnarray}
\laMSb &=& ( 310 \pm 100 ) \, \text{MeV}
\end{eqnarray}
corresponding to
\begin{eqnarray}
\alpha \left( M_Z \right) &=& 0.117  \, \pm 0.006  .
\end{eqnarray}

The situation is the same with respect to the dependence on
$m_c^{\star}$:
The RHS of \rf{cons} depends only weakly on $m_c^{\star}$.
Varying \cite{bbb}
\begin{eqnarray}
m_c^{\star} &=& (1.29 \pm 0.07) \, \text{GeV}
\end{eqnarray}
within the quoted range affects the RHS of \rf{cons} by $3\%$, i.e.\ it
is negligible compared to the uncertainty in $B_K$.  Yet the \kkmd\/
depends on $m_c^{\star}$ sizeably.

For completeness we list the remaining parameters entering the
analysis of $\ek$ \cite{pdg}:
\begin{eqnarray}
&&G_F \; =\;  1.17 \cdot 10^{-5} \, \text{GeV}^{-2},  \quad
F_K \; = \; 161 \, \text{MeV}, \quad
m_K \; = \; 498 \, \text{MeV}, \nonumber \\
&& \Delta \, m_K \; = \;   3.52 \cdot 10^{-15} \,  \text{GeV}, \quad
M_W \; = \; 80.22 \, \text{GeV},  \quad
m_b^{\star} \; = \; 4.2 \, \text{GeV}
\nonumber
\end{eqnarray}
and the measured value for $|\ek|$ has been given in \rf{expek}.
The uncertainties of these quantities are irrelevant for the analysis.

Finally we list the additional input parameters needed for the
\bbm: The \bbmd\/  parameter $x_d=0.78 \pm 0.05$ enters the calculation
in the combination
\begin{eqnarray}
\Delta m_{B_d} &=& x_d/\tau_{B_d}  \; = \;
    ( 0.496 \pm 0.032 )\, ps^{-1}, \label{demb}
\end{eqnarray}
which is the world average presented in \cite{p}.
Yet the largest uncertainty is due to the hadronic parameters
$F_{B_d}$ and $B_{B_d}$ appearing in the form
\begin{eqnarray}
F_{B_d} \sqrt{B_{B_d}} &=& ( 195 \pm 45 ) \, \text{MeV}. \label{fbd}
\end{eqnarray}
This result has been obtained with lattice methods \cite{def} and QCD
sum rules \cite{bpbd}.
The ratio $F_{B_s}/F_{B_d}$ has been well determined from the lattice
\cite{def}:
\begin{eqnarray}
\frac{F_{B_s}}{F_{B_d}} &=& 1.22 \pm 0.04 \label{fbs}.
\end{eqnarray}
Further we will need the meson masses $m_{B_d}=5.28 \, \text{GeV}$ and
$m_{B_s}=5.38 \, \text{GeV}$ and the $B_s$ lifetime
$\tau_{B_s}=(1.53 \pm 0.10) \, ps$ \cite{p}.

\section{Bounds on Standard Model parameters}\label{borderline}
\begin{figure}[htb]
\centerline{
\epsfxsize=14cm
\epsffile{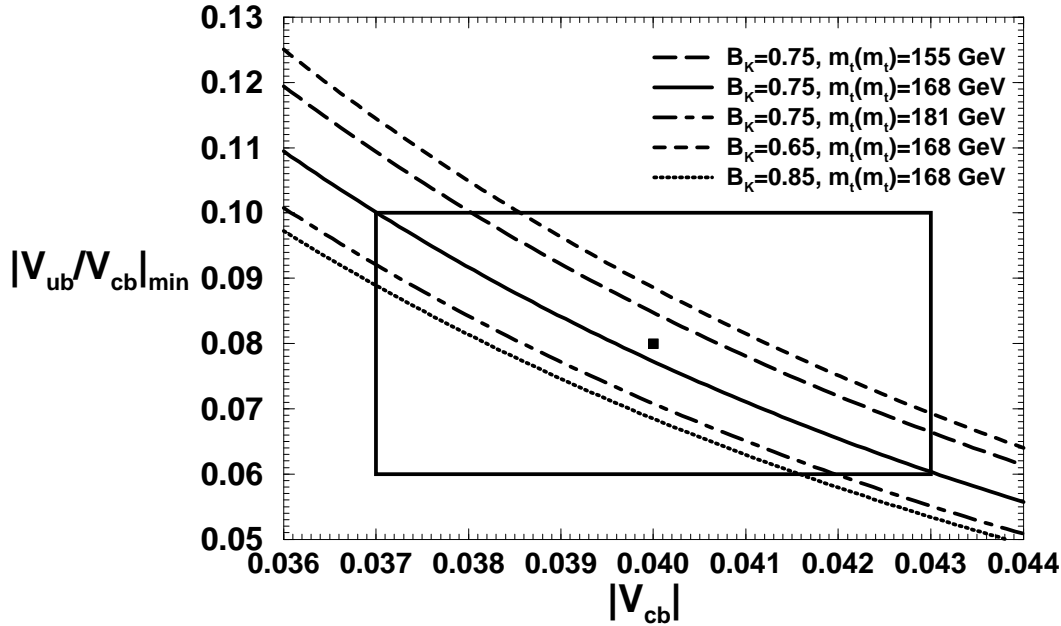}
}
\caption{New physics borderlines for various values of $m_t^{\star}$
and $B_K$. Each pair $(m_t^{\star},B_K)$ defines a curve.  If the
Standard Model is the only source of indirect CP violation in the
neutral Kaon system, the points below the curve are excluded.  The
solid line in the middle corresponds to the central values for
$m_t^{\star}$ and $B_K$ given in sect.~\ref{par}. The rectangle limits
the allowed range for $V_{cb}$ and $|V_{ub}/V_{cb}|$ obtained from
tree-level b-decays according to (\ref{vcb}) and (\ref{vub}). The
point in the middle of the rectangle corresponds to the central values
in (\ref{vcb}) and (\ref{vub}).  }\label{vmin}
\end{figure}

As explained in the previous section the final error bar of the CKM
phase $\delta$ determined from $\ek$ is due to the uncertainties in
$V_{cb}$, $|V_{ub}/V_{cb}|$, $B_K$ and $m_t^{\star}$.  Yet it is
well-known that the unitarity of the CKM Matrix constrains the allowed
range for these four quantities: If one fixes three of them, a lower
bound for the fourth one can be obtained, because otherwise \rf{cons}
yields no real solutions for $\cos \delta$.  In terms of the improved
Wolfenstein parameters \rf{wolf} these solutions appear as the
intersection points of a hyperbola with a circle.  The lower bound
solution corresponds to a set of parameters for which the hyperbola
touches the circle in one point (see \cite{blo} for details).  Prior
to the discovery of the top quark this method was used to find a lower
bound on the top quark mass (see e.g.\ \cite{blo,ajb}).  Now in the
top era it is more useful to determine the allowed region for the
other two fundamental Standard Model parameters in the game, $V_{cb}$
and $|V_{ub}/V_{cb}|$. This is shown in fig.~\ref{vmin}. The ranges
\rf{vcb} and \rf{vub} correspond to a rectangle in fig.~\ref{vmin}.
For each pair $(m_t^{\star},B_K)$ the constraint from $\ek$ defines a
curve in fig.~\ref{vmin} such that only the region above this curve is
allowed.

We emphasize that the central values for the input parameters given in
\rf{vcb} to \rf{bkrange} are close to the borderline curve depicted in
fig.~\ref{vmin}. With the old LO value for $\eta_3$ the central values
would even seem to contradict the measured value for $\ek$.  The
minimal value for $|V_{ub}/V_{cb}|$ equals 0.0778, if the central
values in \rf{vcb}, \rf{mass} and \rf{bkrange} are chosen for the
other parameters. Conversely the minimal values for the other
parameters read $V_{cb, \, \text{min}}=0.397$, $m^{\star}_{t,
\,\text{min}} = 164\, \text{GeV}$ and $B_{K, \, \text{min} }= 0.729$,
if the remaining three ones equal the central values chosen in
sect.~\ref{par}.  Of course varying these parameters to higher values
relaxes the lower bound on the fourth one. Altogether the constraint
from $\ek$ rules out almost one half of the parameter space of
sect.~\ref{par}.

{From} these remarks it is clear that $\ek$ strongly constrains those
extensions of the Standard Model, in which extra CP-violating
interactions {\sl diminish}\/ $|\ek|$, because then the Standard Model
contribution $|\ek^{\text{SM}}|$ must be larger to accommodate for the
measured value of $|\ek|$.  The lower bound can be summarized in the
following approximate formula
\begin{eqnarray}
\frac{V_{cb}}{0.0397} \,
\left( \frac{ |V_{ub}/V_{cb}| }{0.080}        \right)^{0.27}
\left( \frac{m_t^{\star} }{168 \, \text{GeV}}        \right)^{0.31}
\left( \frac{B_K }{0.75 }        \right)^{0.27}
\left( \frac{ 2.27 \cdot 10^{-3} }{| \ek^{ \text{SM} }| }    \right)^{0.27}
&\geq & 1   .  \label{alb}
\end{eqnarray}
$|V_{ub}/V_{cb}|_{\text{min}}$ determined from (\ref{alb}) concides
with the exact solution to $4\%$ accuracy in the parameter range of
sect.~\ref{par}. For $0.039 \leq V_{cb} \leq 0.041$ the agreement is
better than $2\%$.

\rf{alb} displays the sensitivity of our analysis on $V_{cb}$.
Although $V_{cb}$ is known to a much higher accuracy than
$|V_{ub}/V_{cb}|$, its uncertainty contributes roughly as much to the
final error as the one of $|V_{ub}/V_{cb}|$. The situation is similar
in the analysis of sect.~\ref{vtd}.

Finally we remark that in the vicinity of the lower bound values the
determination of the CKM elements is very sensitive to the input
parameters. Because of the required precision one should use the exact
parametrization (\ref{exact}) of the CKM matrix here.

\section{CKM matrix phenomenology}\label{vtd}
In this section we determine various CKM parameters using
$|\varepsilon_K|$ and the unitarity of the CKM matrix and discuss
the constraints following from \bbm.

\subsection{The CKM phase $\delta$}\label{sect:delta}

By solving eqn.~(\ref{cons}) for $\cos\delta$ we calculate the two
solutions for the phase $\delta$ of the CKM matrix.  For the input
parameters defined in sect.~\ref{par} the resulting $\delta$'s have
been compiled into table~\ref{tab:delta}, where the dependence on the
key parameters $m_t^{\star}$, $B_K$, $V_{cb}$ and
$|V_{ub}/V_{cb}|$ is made explicit.  A dash means that there exists no
solution for these parameters, lines which do not contain a solution
at all have been omitted from the table.  This happens  for small
values of the abovementioned input parameters and served to derive the
bounds on these parameters in sect.~\ref{borderline}.

For our central values we observe the two solutions being very close
to the limits derived in sect.~\ref{borderline}.  This leads to very
asymmetric error bars.  Therefore we first give the central values and
the variation of it for all relevant parameters separately.
\begin{eqnarray}
\delta^{\text{low}} &=&
89^{\circ}
\begin{array}{llll}
+13^{\circ} & +13^{\circ} & +13^{\circ} & +13^{\circ} \\
-28^{\circ} & -23^{\circ} & -14^{\circ} & -17^{\circ}
\end{array}
\nonumber \\
\delta^{\text{high}} &=&
116^{\circ}
\begin{array}{llll}
+24^{\circ} & +24^{\circ} & +12^{\circ} & +15^{\circ} \\
-13^{\circ} & -13^{\circ} & -13^{\circ} & -13^{\circ}
\end{array}
\label{vardelta}
\end{eqnarray}
The variations in (\ref{vardelta}) are meant as follows:
The first number in the lower line for $\delta^{\text{low}}$ and
the first number in the upper line for $\delta^{\text{high}}$
are the two solutions
obtained by pushing $V_{cb}$ to its maximal value $V_{cb}=0.043$ while
keeping the other three parameters fixed to their central values
given in sect.~\ref{par}. Conversely the other three numbers
in these lines represent the variation when the same is done
for $|V_{ub}/V_{cb}|$, $m_t^{\star}$ and $B_K$.
In contrast moving the key parameters to lower values makes the two
solutions for $\delta$ approach until they merge, when the varied
parameter reaches its ``lower bound value'' discussed in the preceding
 section.
The variations on the upper line for $\delta^{\text{low}}$ and the
lower line for $\delta^{\text{high}}$ correspond to these values,
which are
$V_{cb,\text{min}}=0.0397$, $|V_{ub}/V_{cb}|_{\text{min}}=0.0778$,
$m^{\star}_{t,\text{min}}=164\gev$ and $B_{K,\text{min}}=0.729$.

We combine the individual variations in (\ref{vardelta}) to
\begin{eqnarray}
\delta^{\text{low}} &=&
89^{\circ}
{+13^{\circ} \atop -43^{\circ}}
\nonumber \\
\delta^{\text{high}} &=&
116^{\circ}
{+39^{\circ} \atop -13^{\circ}}
\label{errdelta}
\end{eqnarray}
The error in the lines stemming from the lower bounds is motivated by
the observation that the value $\delta=103^{\circ}$ for which the two
solutions merge is essentially independent of the input parameters.
The error in the lines emerging from pushing the input parameters to
their maximally allowed values is obtained by adding the four
individual variations of (\ref{vardelta}) in quadrature. This seems
questionable, because the theoretical errors of the input parameters
may be correlated.  Hence we have also determined the error by finding
simply the maximal value for $\delta^{\text{high}}$ and the minimal
value for $\delta^{\text{low}}$ when all input quantities are varied
within the ranges given in sect.~\ref{par}. These extremal values
correspond to the point $(V_{cb}=0.043,|V_{ub}/V_{cb}|=0.10,
m_t^{\star}=181\gev, B_K=0.85)$, because $\delta^{\text{high}}$ and
$\delta^{\text{low}}$ are monotonous functions of all four arguments.
This results in an error which is only slightly larger than the one
cited in (\ref{errdelta}), $-48^{\circ}$ instead of $-43^{\circ}$ in
$\delta^{\text{low}}$ and $+42^{\circ}$ instead of $+39^{\circ}$ in
$\delta^{\text{high}}$.  This is caused by the fact that $\delta$
varies only very slowly in the parameter region far away from the
central values.  $|V_{td}|$ discussed in the following section shows
the same behaviour, which is evident from the plots in
fig.~\ref{fig:vtd-vcb} and fig.~\ref{fig:vtd-mt}.  Hence the error
bars in (\ref{errdelta}) are clearly not too small.

Let us now remark that
in table~\ref{tab:delta} the error resulting from the variation of the
other parameters entering the calculation is not shown.
It amounts to roughly 3--4 degrees.

The discussion of $\delta$ is especially instructive in conjunction
with the unitarity triangle. We will therefore return to $\delta$
in sect.~\ref{sect:tria}, where we will also see that the additional
incorporation of \bbmd\/ yields a tighter upper bound on $\delta$
than the one in  (\ref{errdelta}).

Once we have in this way obtained the phase $\delta$ from the three
angles $s_{12}$, $s_{23}$ and $s_{13}$ or equivalently $V_{us}$,
$V_{cb}$ and $|V_{ub}/V_{cb}|$, we are by use of (\ref{exact}) able to
derive combinations of CKM elements, which are of special
phenomenological interest.

\subsection{$|V_{td}|$}\label{sect:vtd}

$|V_{td}|$ plays an important role for the parameter $x_d$ of \bbmd.
Especially once the \bbms\/ mixing parameter $x_s$ is measured a
theoretically clean determination of $|V_{td}|$ from the ratio
$x_s/x_d$ will be possible.  The comparison of the result with the
determination of $|V_{td}|$ from $\ek$ presented in the following will
be a viable experimental test of the quark mixing sector.

Table~\ref{tab:vtd} shows the value of $|V_{td}|$ as derived from
$\delta$ in table~\ref{tab:delta}.  As usual we  give both
solutions, the smaller one always corresponds to the smaller value of
$\delta$ and vice versa.  As in the case of $\delta$ a dash means that
there exists no solution for the specific set of parameters.  We
find for the central values and the individual variations
\begin{eqnarray}
\left|V_{td}\right|^{\text{low}} \cdot 10^3 &=&
9.3
\begin{array}{llll}
+0.6 & +0.6 & +0.6 & +0.6 \\
-0.9 & -1.3 & -0.7 & -0.9
\end{array}
\nonumber \\
\left|V_{td}\right|^{\text{high}} \cdot 10^3 &=&
10.6
\begin{array}{llll}
+1.7 & +1.5 & +0.5 & +0.6  \\
-0.6 & -0.6 & -0.5 & -0.5
\end{array}
\label{varvtd}
\end{eqnarray}
The upper line of $\left|V_{td}\right|^{\text{low}}$ and the lower
line of $\left|V_{td}\right|^{\text{high}}$ corresponds to
$V_{cb,\text{min}}$, $|V_{ub}/V_{cb}|_{\text{min}}$,
$m_{t,\text{min}}^{\star}$ and $B_{K,\text{min}}$ (see the values in the
paragraph below (\ref{vardelta})), the lower line of
$\left|V_{td}\right|^{\text{low}}$ and the upper line of
$\left|V_{td}\right|^{\text{high}}$ result from putting the input
parameters to their highest allowed value.

In the same way as in the case of $\delta$ in the last section, we
obtained as combined errors
\begin{eqnarray}
\left|V_{td}\right|^{\text{low}} \cdot 10^3 &=&
9.3 { +0.6 \atop -1.9 } \nonumber \\
\left|V_{td}\right|^{\text{high}} \cdot 10^3 &=&
10.6 { +2.4 \atop -0.6 }
\label{errvtd}
\end{eqnarray}
Again the scanning for the extremal values yields an error which is
not much larger than the addition in quadrature: $-2.4$ instead of
$-1.9$ and $+2.9$ instead of $+2.4$ in (\ref{errvtd}). The extremal
values again correspond to the largest values for all input
parameters.  We remark here that we have also used a third way to
estimate the error of $|V_{td}|$: We have scanned the extremal values
for $|V_{td}|$ for those parameters which lie in a 1$\sigma$ ellipsoid
(\ref{sigma}) around the central values. This has yielded the same
error bar as in (\ref{errvtd}). Yet for the determination of the
quantities to be discussed in the following sections this method is
most useful.

\begin{figure}[htb]
\centerline{
\epsffile{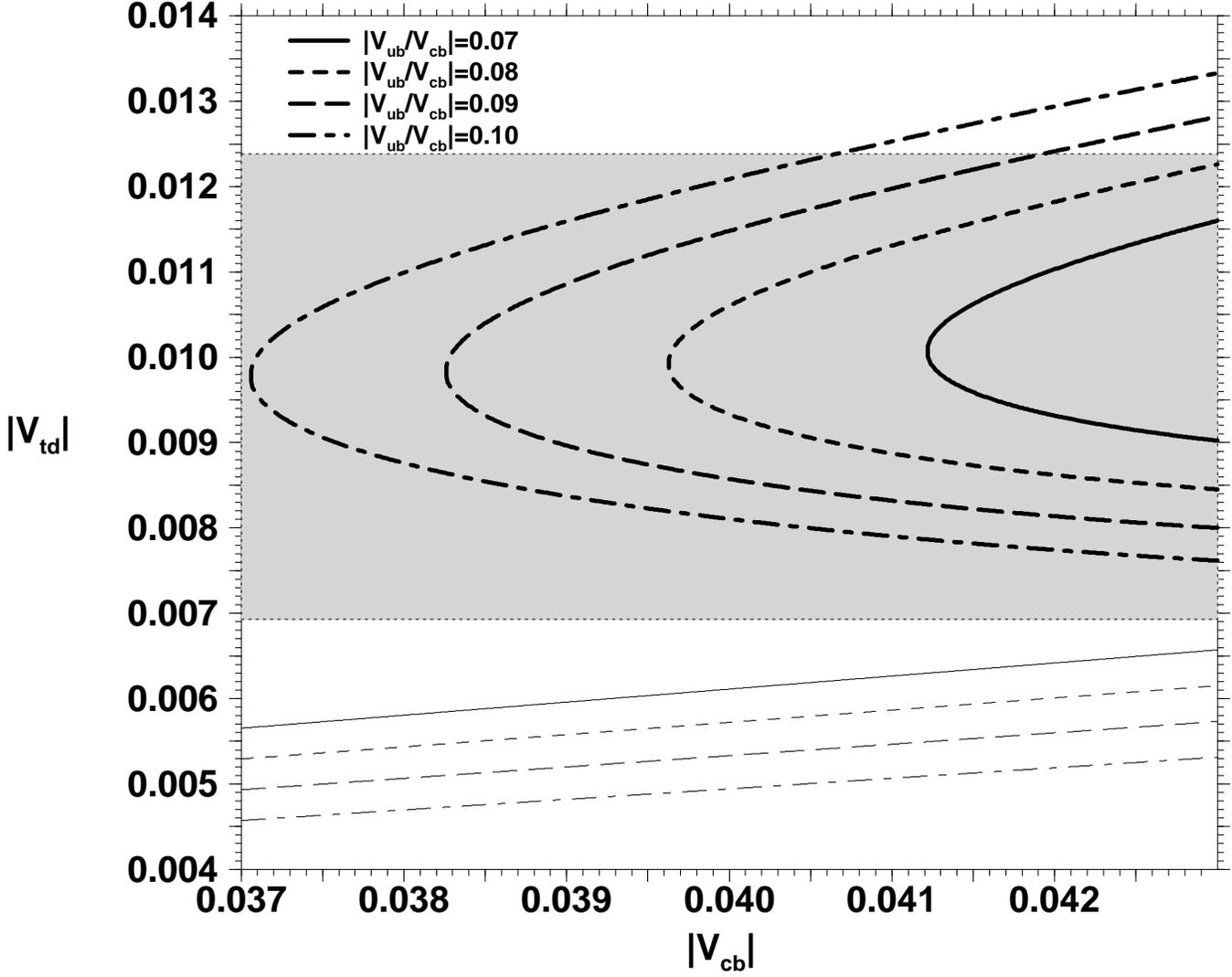}
}
\caption{The dependence of $|V_{td}|$ on $V_{cb}$ for $B_K=0.75$,
$m_t^{\star}=168\protect\gev$ and four values of $|V_{ub}/V_{cb}|$.
The thin lines correspond to $\delta=0$, i.e.\ no CP-violation.  The
shaded area  is  consistent with $x_d$ from
(\protect\ref{vtdbbbar}).}
\label{fig:vtd-vcb}
\end{figure}

Let us discuss the dependence of $|V_{td}|$ on the most important
input parameters in more detail.  In fig.~\ref{fig:vtd-vcb} we plot
the dependence of $|V_{td}|$ on $V_{cb}$ for $|V_{ub}/V_{cb}| = 0.07,
0.08, 0.09, 0.10$ and the other parameters being fixed at their
central values.  For $|V_{ub}/V_{cb}| = 0.06$ we cannot find a
solution.  The curves drawn with thick lines represent the actual
solution for $|V_{td}|$, the thin lines display the value of
$|V_{td}|$, if the phase $\delta$ would be equal to zero.

\begin{figure}[htb]
\centerline{
\epsffile{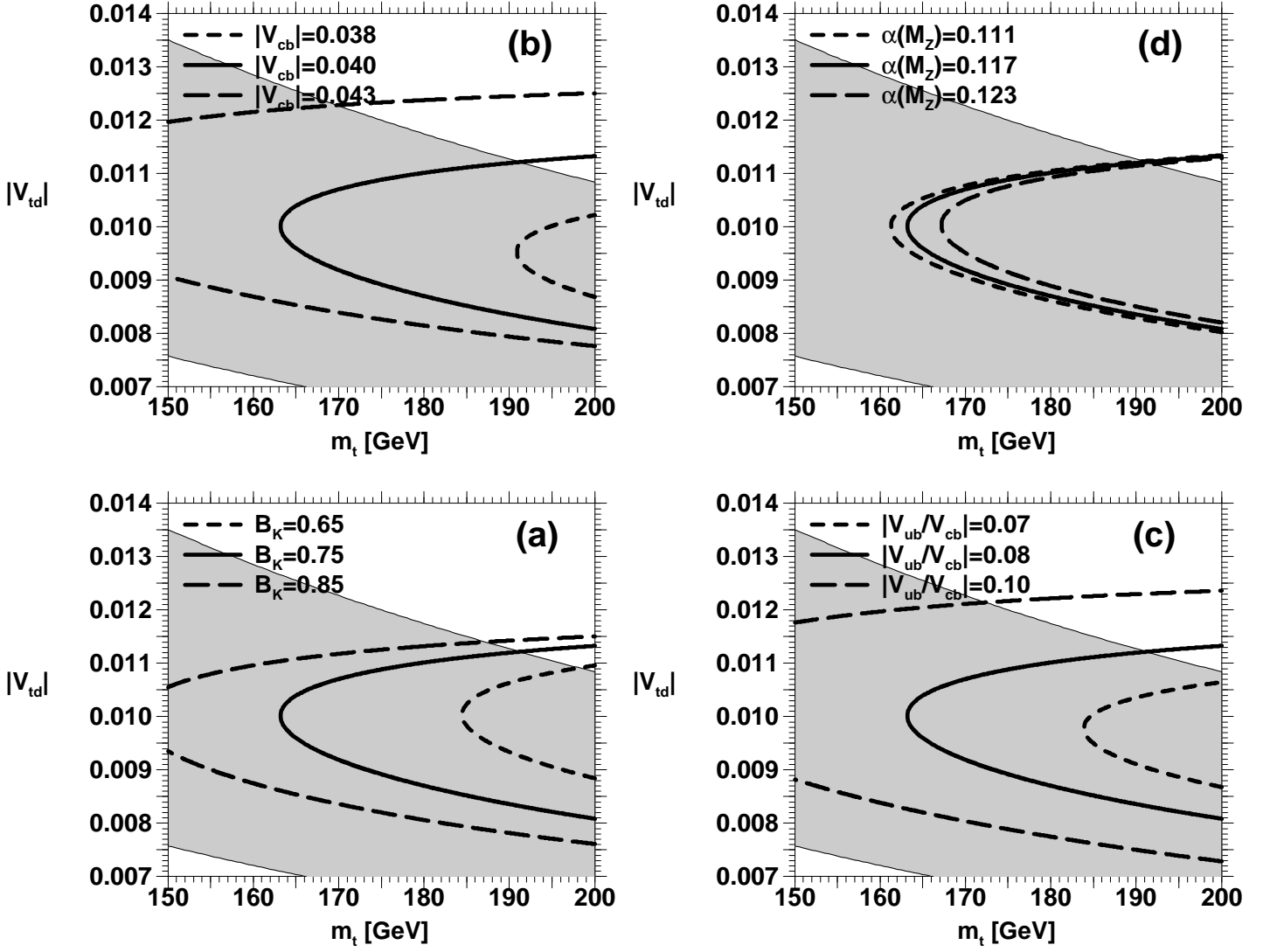}
}
\caption{The dependence of $|V_{td}|$ on $m_t^{\star}$ for three
values of (a) $B_K$, (b) $V_{cb}$, (c) $|V_{ub}/V_{cb}|$ and (d)
$\alpha(M_Z)$.  All other parameters equal their central values of
sect.~\ref{par}.  The shaded area gives the band (\ref{vtdbbbar}) of
$|V_{td}|$'s allowed by the \protect\bbmd\/ parameter $x_d$.  For
large values of the discussed parameters $x_d$ favors the smaller
branch of the solution.}
\label{fig:vtd-mt}
\end{figure}

Let us further compare this to the bound on $|V_{td}|$ which we get
from \bbmd.  The experimentally measured quantities $\Delta m_{B_d}$
and $x_d$ are given by
\begin{eqnarray}
\Delta m_{B_d} \; = \; x_d/\tau_{B_d} &=&
\left| V_{td} \right|^2 \, \left|V_{tb}\right|^2 \,
 \frac{G_F^2}{6\pi^2} \eta_{\text{QCD}} m_B B_{B_d} F_{B_d}^2 M_W^2
S\left(x_t\right) .\label{xd}
\end{eqnarray}
Using $m_t^{\star}=168\gev$ and $\eta_{\text{QCD}}=0.55$ one obtains
with the values of sect.~\ref{par}
\begin{eqnarray}
0.0069 \leq \left|V_{td}\right| \leq 0.0124 .
\label{vtdbbbar}
\end{eqnarray}
This is represented by the shaded band in fig.~\ref{fig:vtd-vcb}.  One
immediately notices, that higher values of $|V_{ub}/V_{cb}|$ and
$V_{cb}$ favor the lower branch of the solution, i.e.\ the smaller
solution for $\delta$.  While for the central values of our analysis
\bbmd\/ implies no additional constraint on $|V_{td}|$, we still get a
tighter upper bound for $|V_{td}|$ compared to the range
(\ref{errvtd}) implying only $|V_{td}| \leq 0.0130$. From
fig.~\ref{fig:vtd-mt} one can easily verify that varying $m_t^{\star}$
does not yield a bound on $|V_{td}|$ different from (\ref{vtdbbbar})
for the combined analysis of $\ek$ and
\bbmd.  Further note that the band derived from $x_d$ clearly shows
$\delta$ being different from zero in the whole range of values for
$|V_{cb}|$.  This is remarkable, because in the Standard Model the
phase $\delta$ is responsible for the CP violation and $x_d$ is a
quantity having nothing to do with the breakdown of this discrete
symmetry.

Let us now explore the $m_t^{\star}$ dependence of $|V_{td}|$, which
is plotted in fig.~\ref{fig:vtd-mt}.  The solid curve is identical for
(a)--(c) and corresponds to the central values of sect.~\ref{par}, we
additionally varied in (a) $B_K = 0.65, 0.75, 0.85$, in (b) $V_{cb} =
0.038, 0.040, 0.043$ and in (c) $|V_{ub}/V_{cb}| = 0.07, 0.08, 0.10$.
No solution was obtained for (b) $V_{cb} = 0.037$ and (c)
$|V_{ub}/V_{cb}| = 0.06$.  The band displayed in grey again shows the
values allowed for $|V_{td}|$ from $x_d$.  Clearly, for larger values
of $m_t^{\star}$, $V_{cb}$ and $|V_{ub}/V_{cb}|$ the constraint from
$x_d$ favors the lower branch of the solution for $|V_{td}|$.

Fig.~\ref{fig:vtd-mt}(d) shows the variation of $|V_{td}|$ vs.\
$m_t^{\star}$ with the value of the strong coupling normalized at
$M_Z$, $\alpha_s\left(M_Z\right)$.  One notices, that the influence of
$\alpha_s\left(M_Z\right)$ far off the point, where the two solutions
merge is quite small.  As one expects, the variation of the value
$m_t$ at the point, where the branches meet is quite large, it amounts
to about 6\gev.  This fact was already discussed at the end of
sect.~\ref{borderline}.

\subsection{Prediction for $x_s$}\label{sectxs}
It is well-known (see e.g.\ \cite{ali}) that an analysis using both
$x_d$ and the \bbms\/ parameter $x_s$ allows for a much more precise
determination of $|V_{td}|$ than the investigation of $x_d$ alone.
The main reason for this is the fact that the hadronic uncertainties
in the ratio $x_d/x_s$ are reduced to SU(3) breaking effects and
are thereby much smaller than in $x_d$ or $x_s$ alone. Further
$|V_{ts}|$ is known very well, because it is related to $V_{cb}$ via
the unitarity of the CKM matrix. The present experimental bound on
$x_s$ does not constrain the ranges (\ref{errvtd}) and (\ref{vtdbbbar})
for $|V_{td}|$ further. Therefore we will instead predict  a range for
$\Delta m_{B_s}$ and $x_s$ from our result (\ref{errvtd}).

We will the mass difference $\Delta m_{B_q}=x_q/ \tau_{B_q}$ with
$q=d,s$ in our formulas. From (\ref{xd}) and the analogous formula for
$x_s$ one finds
\begin{eqnarray}
\Delta m_{B_s} &=& \Delta m_{B_d} \, \frac{|V_{ts}|^2}{|V_{td}|^2}
           \, \frac{1}{\widetilde{R}_{ds}}  \label{su3}
\end{eqnarray}
with
\begin{eqnarray}
\widetilde{R}_{ds}  &=& \frac{m_{B_d}F_{B_d}^2 B_{B_d}
                            }{m_{B_s}F_{B_s}^2 B_{B_s}  }  .
\end{eqnarray}
$\widetilde{R}_{ds} $ equals 1 in the SU(3) limit. The SU(3) breaking
in the decay constants is encoded in (\ref{fbs}). Setting
\begin{eqnarray}
\widetilde{R}_{ds} &=& 0.66 \pm 0.08 \nonumber
\end{eqnarray}
one gets from (\ref{su3})
\begin{eqnarray}
\Delta m_{B_s} &=&  ( 0.76 \pm 0.11  )\, ps^{-1} \cdot
      \frac{|V_{ts}|^2}{|V_{td}|^2} \label{bsnum}  .
\end{eqnarray}
Now for $|V_{td}|=9.3 \cdot 10^{-3}$ one finds $\Delta
m_{B_s}=(13.4\pm 1.9)\, ps^{-1}$ corresponding to $x_s=(20.5 \pm 3.2)$
for $\tau_{B_s}=1.53 \pm 0.10$ \cite{p}.  Equivalently $|V_{td}|=10.6
\cdot 10^{-3}$ yields $\Delta m_{B_s}=(10.2 \pm 1.5)\, ps^{-1}$ and
$x_s=(15.6 \pm 2.4)$. These values are well above the present lower
bound $\Delta m_{B_s} > 6.0 \,ps^{-1}$ from the ALEPH collaboration
\cite{jak}.
In order to find the range for $\Delta m_{B_s}$ consistent with
$\ek$ and $x_d$ in the  parameter range of sect.~\ref{par}
we use two different methods: First we scan the full range
yielding
\begin{eqnarray}
6.3 \, ps^{-1} &\leq& \Delta m_{B_s} \; \leq \; 33 \, ps^{-1} \nonumber  ,
\end{eqnarray}
where the error in (\ref{bsnum}) has been included.
Second we restrict the input parameters to the 1$\sigma$-ellipsoid
\begin{eqnarray}
\left( \frac{ |V_{ub}/V_{cb}| - 0.08 }{ 0.02}   \right)^2  +
\left( \frac{ V_{cb} - 0.040 }{ 0.003}   \right)^2  +
\left( \frac{ m_t^{\star} - 168\, \text{GeV} }{ 13\, \text{GeV}}  \right)^2  +
\left( \frac{ B_K - 0.75 }{ 0.10 }   \right)^2  & \leq & 1, \label{sigma}
\end{eqnarray}
which would be the natural range, if all errors were statistical.
Here we find
\begin{eqnarray}
6.8 \, ps^{-1} &\leq& \Delta m_{B_s} \; \leq \; 24 \, ps^{-1}
\nonumber
\end{eqnarray}
showing that only  the upper bound is sensitive to the border region
of the parameter space.
For our final prediction we use the arithmetic mean of both estimates:
\begin{eqnarray}
6.5 \, ps^{-1} &\leq& \Delta m_{B_s} \; \leq \; 28 \, ps^{-1}
\end{eqnarray}
This corresponds to
\begin{eqnarray}
 9.3     &\leq&  x_{s} \; \leq \; 46   .
\end{eqnarray}
Future stronger bounds on $\Delta m_{B_s}$ may be used to rule out the
higher solution for $|V_{td}|$ in a part of the parameter space: Since
to 1\% accuracy $|V_{ts}|=0.98 V_{cb}$, the relation (\ref{bsnum})
defines a straight line in fig.~\ref{fig:vtd-vcb} excluding the values
for $|V_{td}|$ above this line.

\subsection{$\imag \lambda_t$}

In the discussion of CP violation $\imag \lambda_t$ is of utmost
importance. It is proportional to the Jarlskog parameter,
\begin{eqnarray}
2 J_{\text{CP}} &=& V_{ud} V_{us} \cdot \imag \lambda_t
= \lambda^6 A^2 \bar \eta + O\left(\lambda^8\right),
\end{eqnarray}
and encodes the same experimental information, because the value of
$V_{ud} V_{us}$ is precisely known.  For example
$\varepsilon_{K}'/\ek$ is proportional to $\imag \lambda_t$.  We
tabulate $\imag \lambda_t$ in table~\ref{tab:imlat}.  Here the lower
solution for $\delta$ corresponds to the higher value of $\imag
\lambda_t$ and vice versa.  For our standard choice of parameters from
sect.~\ref{par} we find
\begin{eqnarray}
10^4 \cdot \imag \lambda_t^{\text{low}} &=&
1.15
\begin{array}{llll}
+0.07 & +0.06 & +0.09 & +0.09 \\
-0.20 & -0.12 & -0.15 & -0.19
\end{array}
\nonumber \\
10^4 \cdot \imag \lambda_t^{\text{high}} &=&
1.28
\begin{array}{llll}
+0.03 & +0.19 & +0.00 & +0.00 \\
-0.05 & -0.07 & -0.03 & -0.03
\label{varimlat}
\end{array}
\end{eqnarray}
The upper line of $\imag \lambda_t^{\text{low}}$ and the lower line of
$\imag \lambda_t^{\text{high}}$ corresponds to $V_{cb,\text{min}}$,
$|V_{ub}/V_{cb}|_{\text{min}}$, $m_{t,\text{min}}^{\star}$ and
$B_{K,\text{min}}$ (see the values in the paragraph below
(\ref{vardelta})), the lower line of $\imag \lambda_t^{\text{low}}$
and the upper line of $\imag \lambda_t^{\text{high}}$ result from
putting the input parameters to their highest allowed value.  Note
that $\imag \lambda_t^{\text{high}}$ is not a monotonous function of
the input parameters, for our central values of $m_t^{\star}$ and
$B_K$ we are already close to the maximum.

{From} the analysis of $\ek$ alone we find for a scan of the whole
parameter range the result
\begin{eqnarray}
0.71 \cdot 10^{-4} \leq \imag \lambda_t \leq 1.68 \cdot 10^{-4} .
\label{imlatek}
\end{eqnarray}
Next we include the constraint from $x_d$: We now find the lower bound
in the full parameter range in (\ref{imlatek}) shifted from 0.71 to
0.81.  For the parameter range (\ref{sigma}) we find
\begin{eqnarray}
0.89 \cdot 10^{-4} \leq \imag \lambda_t \leq 1.51 \cdot 10^{-4} .
\end{eqnarray}
We combine the two estimates to our final result
\begin{eqnarray}
0.85 \cdot 10^{-4} \leq \imag \lambda_t \leq 1.60 \cdot 10^{-4} .
\end{eqnarray}

\begin{figure}[htb]
\centerline{
\epsffile{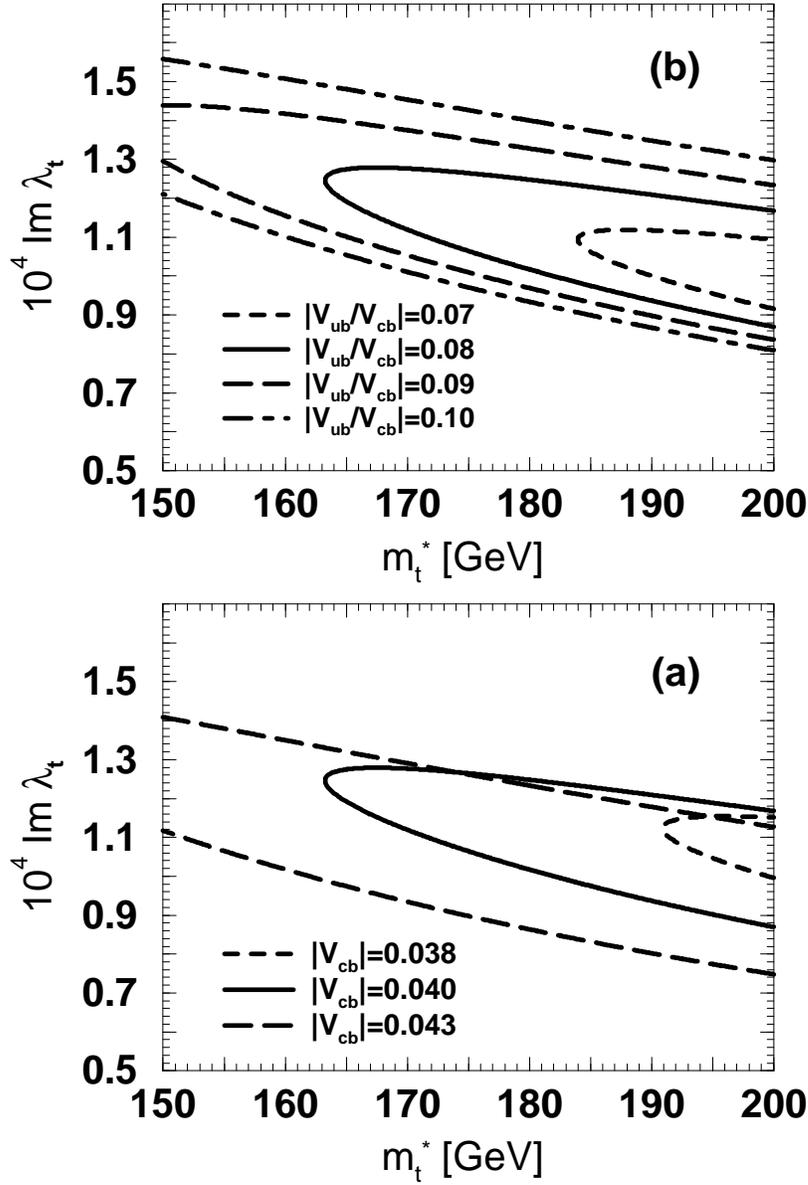}
}
\caption{The dependence of $\protect\imag \lambda_t$ on $m_t^{\star}$ for
(a) three values of $|V_{cb}|$ and (b) four values of
$|V_{ub}/V_{cb}|$.  In plot (a) one observes that the higher solution
for $\protect\imag \lambda_t$  is stable with respect
to the variation of $|V_{cb}|$, whereas the lower branch depends
quite strongly on this parameter.}
\label{fig:imlat-mt}
\end{figure}

The $m_t^{\star}$ dependence of $\imag \lambda_t$ may be looked at in
fig.~\ref{fig:imlat-mt}.  Plot (a) shows this dependence for three
values of $|V_{cb}|$, plot (b) uses four values for $|V_{ub}/V_{cb}|$.
Note that the result for $\imag \lambda_t$ on the upper branch is
essentially independent of $V_{cb}$, whereas the lower branch varies
quite strongly with $V_{cb}$.

\subsection{$\bar\rho$, $\bar\eta$ and the unitarity triangle}\label{sect:tria}
Our knowledge about the CKM parameters related to CP violation is
usually expressed by the unitarity triangle introduced in
sect.~\ref{sectckm}.

Using $\delta$ from tab.~\ref{tab:delta}, one obtains the allowed
pairs of $(\bar\rho,\bar\eta)$ listed in the tables~\ref{tab:rhobar},
\ref{tab:etabar}.  Note that this table is constructed solely from
the unitarity of the CKM matrix and the constraint from $|\ek|$.
The additional constraint from $x_d$ can be included by recalling
from (\ref{rhobar}) that
\begin{eqnarray}
\left( 1- \bar \rho    \right)^2 +  \bar \eta ^2 &=&
\left| \frac{ V_{td} V_{tb}}{ V_{cd} V_{cb}} \right|^2 .
\label{circle}
\end{eqnarray}
Since to $0.2\%$ accuracy $|V_{cd}|=V_{us}=0.22$ and $|V_{tb}|=1$ the
determination of $|V_{td}|$ from (\ref{xd}) yields a circle in the
$\bar \rho$-$\bar \eta$-plane around $(1,0)$ for each pair
$(m_t^{\star}, V_{cb})$.

\begin{figure}[htb]
\centerline{
\epsffile{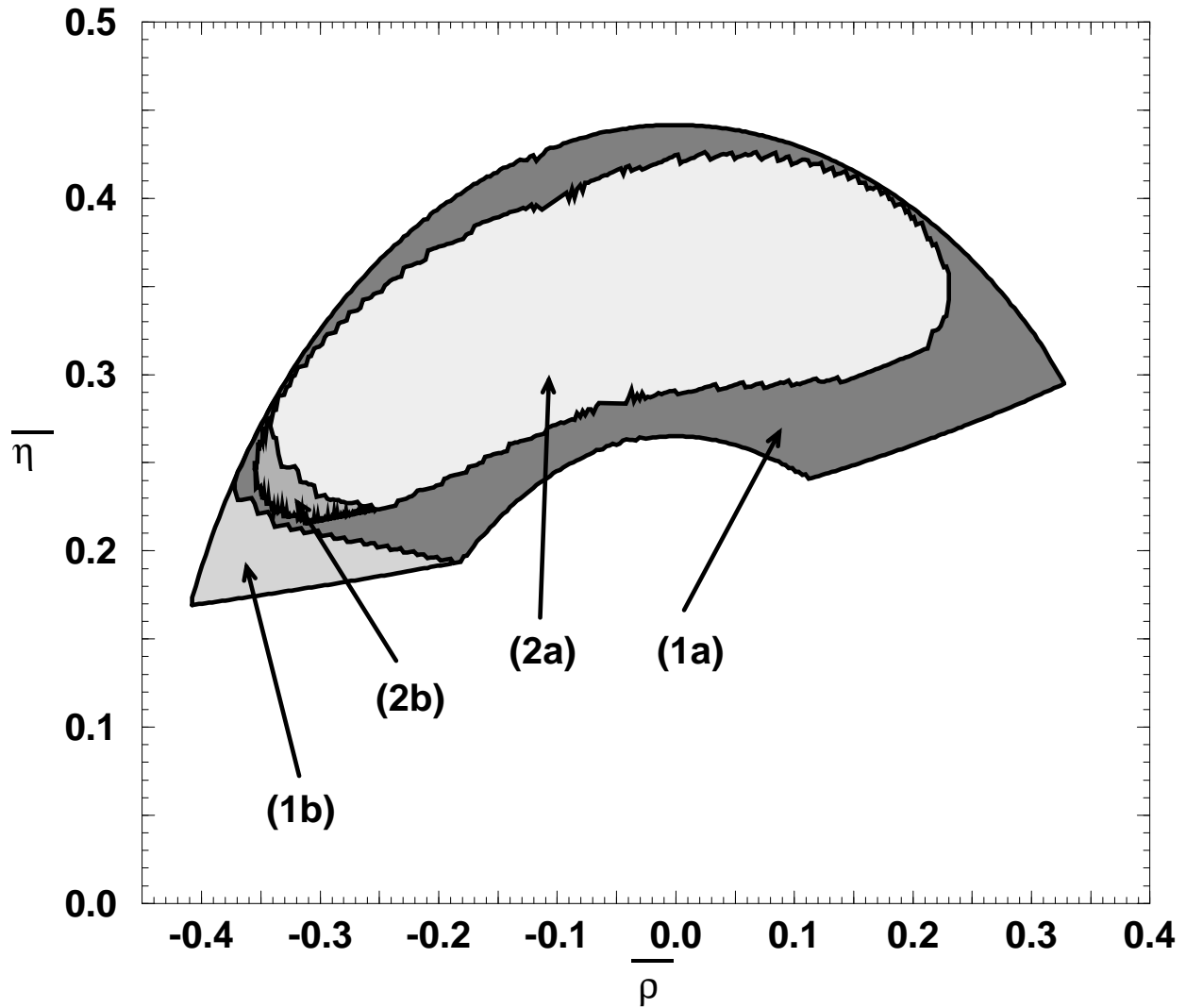}
}
\caption{The allowed region for the pair $(\bar\rho,\bar\eta)$
consistent with $\protect\ek$ and $x_d$.  Area (1a) is obtained from
a scan over the full parameter range of sect.~\protect\ref{par}.
Region (2a) corresponds to the parameters in the $1\sigma$ ellipsoid
(\protect\ref{sigma}).  Areas (1b) and (2b) are consistent with
$\protect\ek$, but not with $x_d$.}
\label{fig:ut1}
\end{figure}

In fig.~\ref{fig:ut1} we display the allowed region for the pair
$(\bar\rho,\bar\eta)$ including the constraint from $x_d$ (\ref{xd})
described in sect.~\ref{sect:vtd}.  Applying this constraint results
in cutting the allowed region of $(\bar\rho,\bar\eta)$ on the left
side of the figure.  To obtain a reasonable estimate of the error
present in the analysis, we have again used two methods.  The area
displayed in dark grey results from varying the input parameters
$B_K$, $m_t^{\star}$, $V_{cb}$, $|V_{ub}/V_{cb}|$ in the full parameter
range described in sect.~\ref{par}, the area displayed in light grey
is obtained by requiring the used parameters to lie within the four
dimensional $1\sigma$-ellipsoid described in (\ref{sigma}).

{From} fig.~\ref{fig:ut1} we read off the following allowed regions for
$(\bar\rho,\bar\eta)$ and the angles $\alpha$, $\beta$, $\gamma$ in
fig.~\ref{unipict}:
\begin{eqnarray}
\begin{array}
	{r@{\,\leq\,}c@{\,\leq\,}l@{\hspace{1cm}}r@{\,\leq\,}c@{\,\leq\,}l}
-0.37 & \bar\rho & 0.33 & -0.34 & \bar\rho & 0.23 \\
0.19 & \bar\eta & 0.44 & 0.22 & \bar\eta & 0.43 \\
22.3^{\circ} &\alpha& 114.3^{\circ} & 26.2^{\circ} &\alpha& 102.1^{\circ} \\
9.1^{\circ} &\beta& 26.2^{\circ} & 14.9^{\circ} &\beta& 26.2^{\circ} \\
42.0^{\circ} &\gamma& 148.2^{\circ} & 55.5^{\circ} &\gamma& 143.3^{\circ}
\end{array}
\end{eqnarray}
The ranges quoted in the first column correspond to the error estimate
by the box-scan, the second column to the $1\sigma$-ellipsoid method.
Again we quote as our final range the arithmetic mean of both estimates:
\begin{eqnarray}
\begin{array}{r@{\,\leq\,}c@{\,\leq\,}l}
-0.36 & \bar\rho & 0.28 \\
0.21 & \bar\eta & 0.44 \\
24^{\circ} &\alpha& 108 ^{\circ} \\
12^{\circ} &\beta& 26 ^{\circ} \\
49^{\circ} &\gamma& 146^{\circ}
\end{array}
\end{eqnarray}
CP asymmetries in the B  system are proportional to the sines of
$2 \alpha$, $2 \beta$ or $2 \gamma$. We can only reliably predict
$\sin 2 \beta$:
\begin{eqnarray}
0.41 \leq \sin 2 \beta \leq 0.79 \nonumber  ,
\end{eqnarray}
where the upper bound stems solely from $|V_{ub}/V_{cb}|\leq 0.10$
(see \cite{blo}).

Since to $0.1\%$ accuracy $\delta=\gamma$ we can now improve the range
(\ref{errdelta}) by the inclusion of the constraint from $x_d$:
\begin{eqnarray}
49^{\circ}  \leq  \delta \leq  146^{\circ} \nonumber .
\end{eqnarray}

\begin{table}[htb]
\squeezetable
\begin{tabular}[tb]{@{\hspace{3pt}}c@{\hspace{3pt}}c@{\hspace{3pt}}c@{\hspace{6pt}}r@{\hspace{3pt}}r@{\hspace{6pt}}r@{\hspace{3pt}}r@{\hspace{6pt}}r@{\hspace{3pt}}r@{\hspace{6pt}}r@{\hspace{3pt}}r@{\hspace{6pt}}r@{\hspace{3pt}}r@{\hspace{6pt}}r@{\hspace{3pt}}r@{\hspace{6pt}}r@{\hspace{3pt}}r}
\multicolumn{3}{c}{$V_{cb}$}&\multicolumn{2}{c}{0.037}&\multicolumn{2}{c}{0.038}&\multicolumn{2}{c}{0.039}&\multicolumn{2}{c}{0.040}&\multicolumn{2}{c}{0.041}&\multicolumn{2}{c}{0.042}&\multicolumn{2}{c}{0.043}\\
\hline
$m_t^*$&$B_K$&$\left|\frac{V_{ub}}{V_{cb}}\right|$&\multicolumn{14}{c}{$\delta$}\\\hline\hline
155&0.65&0.08&\multicolumn{2}{c}{---}&\multicolumn{2}{c}{---}&\multicolumn{2}{c}{---}&\multicolumn{2}{c}{---}&\multicolumn{2}{c}{---}&\multicolumn{2}{c}{---}&86&120\\
155&0.65&0.09&\multicolumn{2}{c}{---}&\multicolumn{2}{c}{---}&\multicolumn{2}{c}{---}&\multicolumn{2}{c}{---}&98&110&81&126&72&133\\
155&0.65&0.10&\multicolumn{2}{c}{---}&\multicolumn{2}{c}{---}&\multicolumn{2}{c}{---}&92&118&79&129&71&136&65&141\\
\hline
155&0.75&0.07&\multicolumn{2}{c}{---}&\multicolumn{2}{c}{---}&\multicolumn{2}{c}{---}&\multicolumn{2}{c}{---}&\multicolumn{2}{c}{---}&\multicolumn{2}{c}{---}&83&119\\
155&0.75&0.08&\multicolumn{2}{c}{---}&\multicolumn{2}{c}{---}&\multicolumn{2}{c}{---}&\multicolumn{2}{c}{---}&90&115&77&127&69&134\\
155&0.75&0.09&\multicolumn{2}{c}{---}&\multicolumn{2}{c}{---}&\multicolumn{2}{c}{---}&85&122&75&131&67&137&61&142\\
155&0.75&0.10&\multicolumn{2}{c}{---}&\multicolumn{2}{c}{---}&83&126&74&134&67&140&61&144&56&148\\
\hline
155&0.85&0.06&\multicolumn{2}{c}{---}&\multicolumn{2}{c}{---}&\multicolumn{2}{c}{---}&\multicolumn{2}{c}{---}&\multicolumn{2}{c}{---}&\multicolumn{2}{c}{---}&89&111\\
155&0.85&0.07&\multicolumn{2}{c}{---}&\multicolumn{2}{c}{---}&\multicolumn{2}{c}{---}&\multicolumn{2}{c}{---}&93&110&77&124&68&132\\
155&0.85&0.08&\multicolumn{2}{c}{---}&\multicolumn{2}{c}{---}&\multicolumn{2}{c}{---}&83&121&73&130&65&137&59&141\\
155&0.85&0.09&\multicolumn{2}{c}{---}&96&111&80&126&71&134&64&140&58&144&53&148\\
155&0.85&0.10&92&117&79&129&70&136&64&142&58&146&53&149&49&152\\
\hline
\hline
168&0.65&0.07&\multicolumn{2}{c}{---}&\multicolumn{2}{c}{---}&\multicolumn{2}{c}{---}&\multicolumn{2}{c}{---}&\multicolumn{2}{c}{---}&\multicolumn{2}{c}{---}&95&109\\
168&0.65&0.08&\multicolumn{2}{c}{---}&\multicolumn{2}{c}{---}&\multicolumn{2}{c}{---}&\multicolumn{2}{c}{---}&\multicolumn{2}{c}{---}&84&121&74&130\\
168&0.65&0.09&\multicolumn{2}{c}{---}&\multicolumn{2}{c}{---}&\multicolumn{2}{c}{---}&96&113&80&127&72&134&65&140\\
168&0.65&0.10&\multicolumn{2}{c}{---}&\multicolumn{2}{c}{---}&91&119&79&130&71&137&65&142&59&146\\
\hline
168&0.75&0.07&\multicolumn{2}{c}{---}&\multicolumn{2}{c}{---}&\multicolumn{2}{c}{---}&\multicolumn{2}{c}{---}&\multicolumn{2}{c}{---}&82&121&72&130\\
168&0.75&0.08&\multicolumn{2}{c}{---}&\multicolumn{2}{c}{---}&\multicolumn{2}{c}{---}&89&116&76&128&68&135&62&140\\
168&0.75&0.09&\multicolumn{2}{c}{---}&\multicolumn{2}{c}{---}&85&123&74&132&67&138&61&143&56&147\\
168&0.75&0.10&\multicolumn{2}{c}{---}&83&126&73&134&66&140&61&144&56&148&51&151\\
\hline
168&0.85&0.06&\multicolumn{2}{c}{---}&\multicolumn{2}{c}{---}&\multicolumn{2}{c}{---}&\multicolumn{2}{c}{---}&\multicolumn{2}{c}{---}&86&114&74&126\\
168&0.85&0.07&\multicolumn{2}{c}{---}&\multicolumn{2}{c}{---}&\multicolumn{2}{c}{---}&91&112&76&125&68&133&61&138\\
168&0.85&0.08&\multicolumn{2}{c}{---}&\multicolumn{2}{c}{---}&83&122&72&131&65&137&59&142&54&146\\
168&0.85&0.09&97&111&80&127&71&134&64&140&58&144&53&148&49&151\\
168&0.85&0.10&79&129&70&136&64&142&58&146&53&150&49&153&45&155\\
\hline
\hline
181&0.65&0.07&\multicolumn{2}{c}{---}&\multicolumn{2}{c}{---}&\multicolumn{2}{c}{---}&\multicolumn{2}{c}{---}&\multicolumn{2}{c}{---}&94&111&78&125\\
181&0.65&0.08&\multicolumn{2}{c}{---}&\multicolumn{2}{c}{---}&\multicolumn{2}{c}{---}&\multicolumn{2}{c}{---}&84&122&74&131&66&137\\
181&0.65&0.09&\multicolumn{2}{c}{---}&\multicolumn{2}{c}{---}&97&112&81&127&72&135&65&140&59&144\\
181&0.65&0.10&\multicolumn{2}{c}{---}&93&118&80&130&71&137&65&142&59&146&54&150\\
\hline
181&0.75&0.06&\multicolumn{2}{c}{---}&\multicolumn{2}{c}{---}&\multicolumn{2}{c}{---}&\multicolumn{2}{c}{---}&\multicolumn{2}{c}{---}&\multicolumn{2}{c}{---}&78&122\\
181&0.75&0.07&\multicolumn{2}{c}{---}&\multicolumn{2}{c}{---}&\multicolumn{2}{c}{---}&\multicolumn{2}{c}{---}&82&121&71&130&64&136\\
181&0.75&0.08&\multicolumn{2}{c}{---}&\multicolumn{2}{c}{---}&90&116&77&128&68&135&62&140&56&145\\
181&0.75&0.09&\multicolumn{2}{c}{---}&86&122&75&132&67&138&61&143&55&147&51&150\\
181&0.75&0.10&84&125&74&134&67&140&61&145&56&148&51&151&47&154\\
\hline
181&0.85&0.06&\multicolumn{2}{c}{---}&\multicolumn{2}{c}{---}&\multicolumn{2}{c}{---}&\multicolumn{2}{c}{---}&86&115&73&126&65&133\\
181&0.85&0.07&\multicolumn{2}{c}{---}&\multicolumn{2}{c}{---}&92&111&77&126&68&133&61&139&55&143\\
181&0.85&0.08&\multicolumn{2}{c}{---}&84&121&73&131&65&137&59&142&54&146&49&150\\
181&0.85&0.09&81&126&71&134&64&140&58&145&53&148&49&151&45&154\\
181&0.85&0.10&71&136&64&142&58&146&53&150&49&153&45&155&42&158\\
\end{tabular}

\caption{The two solutions for the phase $\delta$ of the CKM matrix in
degrees as a function of $m_t^{\star}$, $B_K$, $|V_{ub}/V_{cb}|$ and
$|V_{cb}|$.  A dash means no solution.  Lines with no solutions at all
have been omitted.  Values for other input parameters may be
calculated by linear interpolation.}
\label{tab:delta}
\end{table}

\begin{table}[htb]
\squeezetable
\begin{tabular}[tb]{@{\hspace{3pt}}c@{\hspace{3pt}}c@{\hspace{3pt}}c@{\hspace{6pt}}r@{\hspace{3pt}}r@{\hspace{6pt}}r@{\hspace{3pt}}r@{\hspace{6pt}}r@{\hspace{3pt}}r@{\hspace{6pt}}r@{\hspace{3pt}}r@{\hspace{6pt}}r@{\hspace{3pt}}r@{\hspace{6pt}}r@{\hspace{3pt}}r@{\hspace{6pt}}r@{\hspace{3pt}}r}
\multicolumn{3}{c}{$V_{cb}$}&\multicolumn{2}{c}{0.037}&\multicolumn{2}{c}{0.038}&\multicolumn{2}{c}{0.039}&\multicolumn{2}{c}{0.040}&\multicolumn{2}{c}{0.041}&\multicolumn{2}{c}{0.042}&\multicolumn{2}{c}{0.043}\\
\hline
$m_t^*$&$B_K$&$\left|\frac{V_{ub}}{V_{cb}}\right|$&\multicolumn{14}{c}{$\left|V_{td}\right|\cdot10^3$}\\\hline\hline
155&0.65&0.08&\multicolumn{2}{c}{---}&\multicolumn{2}{c}{---}&\multicolumn{2}{c}{---}&\multicolumn{2}{c}{---}&\multicolumn{2}{c}{---}&\multicolumn{2}{c}{---}&9.8&11.5\\
155&0.65&0.09&\multicolumn{2}{c}{---}&\multicolumn{2}{c}{---}&\multicolumn{2}{c}{---}&\multicolumn{2}{c}{---}&10.2&10.9&9.5&11.8&9.1&12.4\\
155&0.65&0.10&\multicolumn{2}{c}{---}&\multicolumn{2}{c}{---}&\multicolumn{2}{c}{---}&9.8&11.2&9.2&12.0&8.9&12.6&8.6&13.0\\
\hline
155&0.75&0.07&\multicolumn{2}{c}{---}&\multicolumn{2}{c}{---}&\multicolumn{2}{c}{---}&\multicolumn{2}{c}{---}&\multicolumn{2}{c}{---}&\multicolumn{2}{c}{---}&9.6&11.2\\
155&0.75&0.08&\multicolumn{2}{c}{---}&\multicolumn{2}{c}{---}&\multicolumn{2}{c}{---}&\multicolumn{2}{c}{---}&9.6&10.8&9.1&11.5&8.9&12.1\\
155&0.75&0.09&\multicolumn{2}{c}{---}&\multicolumn{2}{c}{---}&\multicolumn{2}{c}{---}&9.2&11.1&8.8&11.7&8.6&12.2&8.4&12.7\\
155&0.75&0.10&\multicolumn{2}{c}{---}&\multicolumn{2}{c}{---}&9.0&11.3&8.6&11.9&8.3&12.4&8.1&12.8&8.0&13.2\\
\hline
155&0.85&0.06&\multicolumn{2}{c}{---}&\multicolumn{2}{c}{---}&\multicolumn{2}{c}{---}&\multicolumn{2}{c}{---}&\multicolumn{2}{c}{---}&\multicolumn{2}{c}{---}&9.8&10.7\\
155&0.85&0.07&\multicolumn{2}{c}{---}&\multicolumn{2}{c}{---}&\multicolumn{2}{c}{---}&\multicolumn{2}{c}{---}&9.6&10.4&9.1&11.2&8.9&11.7\\
155&0.85&0.08&\multicolumn{2}{c}{---}&\multicolumn{2}{c}{---}&\multicolumn{2}{c}{---}&9.0&10.8&8.7&11.4&8.5&11.9&8.3&12.3\\
155&0.85&0.09&\multicolumn{2}{c}{---}&9.4&10.1&8.7&11.0&8.4&11.6&8.1&12.0&8.0&12.5&7.9&12.8\\
155&0.85&0.10&9.0&10.3&8.5&11.1&8.2&11.7&7.9&12.1&7.7&12.6&7.6&13.0&7.5&13.4\\
\hline
\hline
168&0.65&0.07&\multicolumn{2}{c}{---}&\multicolumn{2}{c}{---}&\multicolumn{2}{c}{---}&\multicolumn{2}{c}{---}&\multicolumn{2}{c}{---}&\multicolumn{2}{c}{---}&10.2&10.8\\
168&0.65&0.08&\multicolumn{2}{c}{---}&\multicolumn{2}{c}{---}&\multicolumn{2}{c}{---}&\multicolumn{2}{c}{---}&\multicolumn{2}{c}{---}&9.5&11.3&9.2&12.0\\
168&0.65&0.09&\multicolumn{2}{c}{---}&\multicolumn{2}{c}{---}&\multicolumn{2}{c}{---}&9.8&10.7&9.2&11.6&8.8&12.1&8.6&12.6\\
168&0.65&0.10&\multicolumn{2}{c}{---}&\multicolumn{2}{c}{---}&9.5&11.0&9.0&11.7&8.6&12.3&8.4&12.8&8.2&13.2\\
\hline
168&0.75&0.07&\multicolumn{2}{c}{---}&\multicolumn{2}{c}{---}&\multicolumn{2}{c}{---}&\multicolumn{2}{c}{---}&\multicolumn{2}{c}{---}&9.3&11.0&9.0&11.6\\
168&0.75&0.08&\multicolumn{2}{c}{---}&\multicolumn{2}{c}{---}&\multicolumn{2}{c}{---}&9.3&10.6&8.9&11.3&8.6&11.8&8.5&12.3\\
168&0.75&0.09&\multicolumn{2}{c}{---}&\multicolumn{2}{c}{---}&9.0&10.9&8.6&11.5&8.3&12.0&8.1&12.4&8.0&12.8\\
168&0.75&0.10&\multicolumn{2}{c}{---}&8.8&11.0&8.4&11.6&8.1&12.1&7.9&12.5&7.7&12.9&7.6&13.3\\
\hline
168&0.85&0.06&\multicolumn{2}{c}{---}&\multicolumn{2}{c}{---}&\multicolumn{2}{c}{---}&\multicolumn{2}{c}{---}&\multicolumn{2}{c}{---}&9.4&10.5&9.1&11.2\\
168&0.85&0.07&\multicolumn{2}{c}{---}&\multicolumn{2}{c}{---}&\multicolumn{2}{c}{---}&9.3&10.2&8.8&10.9&8.6&11.4&8.5&11.9\\
168&0.85&0.08&\multicolumn{2}{c}{---}&\multicolumn{2}{c}{---}&8.8&10.6&8.4&11.1&8.2&11.6&8.1&12.0&8.0&12.4\\
168&0.85&0.09&9.2&9.8&8.5&10.7&8.1&11.3&7.9&11.7&7.8&12.2&7.7&12.6&7.6&12.9\\
168&0.85&0.10&8.3&10.8&8.0&11.4&7.7&11.8&7.5&12.3&7.4&12.7&7.3&13.1&7.2&13.4\\
\hline
\hline
181&0.65&0.07&\multicolumn{2}{c}{---}&\multicolumn{2}{c}{---}&\multicolumn{2}{c}{---}&\multicolumn{2}{c}{---}&\multicolumn{2}{c}{---}&9.9&10.6&9.3&11.4\\
181&0.65&0.08&\multicolumn{2}{c}{---}&\multicolumn{2}{c}{---}&\multicolumn{2}{c}{---}&\multicolumn{2}{c}{---}&9.3&11.1&8.9&11.7&8.7&12.2\\
181&0.65&0.09&\multicolumn{2}{c}{---}&\multicolumn{2}{c}{---}&9.6&10.4&9.0&11.3&8.6&11.9&8.4&12.3&8.2&12.8\\
181&0.65&0.10&\multicolumn{2}{c}{---}&9.3&10.7&8.8&11.4&8.4&12.0&8.2&12.5&8.0&12.9&7.8&13.3\\
\hline
181&0.75&0.06&\multicolumn{2}{c}{---}&\multicolumn{2}{c}{---}&\multicolumn{2}{c}{---}&\multicolumn{2}{c}{---}&\multicolumn{2}{c}{---}&\multicolumn{2}{c}{---}&9.3&11.0\\
181&0.75&0.07&\multicolumn{2}{c}{---}&\multicolumn{2}{c}{---}&\multicolumn{2}{c}{---}&\multicolumn{2}{c}{---}&9.1&10.8&8.8&11.3&8.6&11.8\\
181&0.75&0.08&\multicolumn{2}{c}{---}&\multicolumn{2}{c}{---}&9.1&10.3&8.7&11.0&8.4&11.5&8.2&12.0&8.1&12.4\\
181&0.75&0.09&\multicolumn{2}{c}{---}&8.8&10.6&8.4&11.2&8.1&11.7&7.9&12.1&7.8&12.5&7.7&12.9\\
181&0.75&0.10&8.6&10.7&8.2&11.3&7.9&11.8&7.7&12.2&7.6&12.6&7.4&13.0&7.3&13.4\\
\hline
181&0.85&0.06&\multicolumn{2}{c}{---}&\multicolumn{2}{c}{---}&\multicolumn{2}{c}{---}&\multicolumn{2}{c}{---}&9.2&10.3&8.9&10.9&8.7&11.4\\
181&0.85&0.07&\multicolumn{2}{c}{---}&\multicolumn{2}{c}{---}&9.1&9.9&8.6&10.7&8.4&11.2&8.3&11.6&8.2&12.0\\
181&0.85&0.08&\multicolumn{2}{c}{---}&8.6&10.3&8.2&10.9&8.0&11.3&7.9&11.8&7.8&12.1&7.7&12.5\\
181&0.85&0.09&8.3&10.4&8.0&11.0&7.8&11.5&7.6&11.9&7.5&12.3&7.4&12.6&7.3&13.0\\
181&0.85&0.10&7.8&11.1&7.6&11.5&7.4&12.0&7.2&12.4&7.1&12.7&7.0&13.1&7.0&13.5\\
\end{tabular}

\caption{The values of $|V_{td}|$ corresponding to the two values of
$\delta$ in table~\protect\ref{tab:delta}.}
\label{tab:vtd}
\end{table}

\begin{table}[htb]
\squeezetable
\begin{tabular}[tb]{@{\hspace{3pt}}c@{\hspace{3pt}}c@{\hspace{3pt}}c@{\hspace{6pt}}r@{\hspace{3pt}}r@{\hspace{6pt}}r@{\hspace{3pt}}r@{\hspace{6pt}}r@{\hspace{3pt}}r@{\hspace{6pt}}r@{\hspace{3pt}}r@{\hspace{6pt}}r@{\hspace{3pt}}r@{\hspace{6pt}}r@{\hspace{3pt}}r@{\hspace{6pt}}r@{\hspace{3pt}}r}
\multicolumn{3}{c}{$V_{cb}$}&\multicolumn{2}{c}{0.037}&\multicolumn{2}{c}{0.038}&\multicolumn{2}{c}{0.039}&\multicolumn{2}{c}{0.040}&\multicolumn{2}{c}{0.041}&\multicolumn{2}{c}{0.042}&\multicolumn{2}{c}{0.043}\\
\hline
$m_t^*$&$B_K$&$\left|\frac{V_{ub}}{V_{cb}}\right|$&\multicolumn{14}{c}{$\mbox{Im}\lambda_t\cdot10^4$}\\\hline\hline
155&0.65&0.08&\multicolumn{2}{c}{---}&\multicolumn{2}{c}{---}&\multicolumn{2}{c}{---}&\multicolumn{2}{c}{---}&\multicolumn{2}{c}{---}&\multicolumn{2}{c}{---}&1.47&1.28\\
155&0.65&0.09&\multicolumn{2}{c}{---}&\multicolumn{2}{c}{---}&\multicolumn{2}{c}{---}&\multicolumn{2}{c}{---}&1.50&1.42&1.57&1.29&1.59&1.21\\
155&0.65&0.10&\multicolumn{2}{c}{---}&\multicolumn{2}{c}{---}&\multicolumn{2}{c}{---}&1.60&1.41&1.65&1.30&1.67&1.22&1.68&1.16\\
\hline
155&0.75&0.07&\multicolumn{2}{c}{---}&\multicolumn{2}{c}{---}&\multicolumn{2}{c}{---}&\multicolumn{2}{c}{---}&\multicolumn{2}{c}{---}&\multicolumn{2}{c}{---}&1.28&1.13\\
155&0.75&0.08&\multicolumn{2}{c}{---}&\multicolumn{2}{c}{---}&\multicolumn{2}{c}{---}&\multicolumn{2}{c}{---}&1.34&1.22&1.38&1.13&1.38&1.07\\
155&0.75&0.09&\multicolumn{2}{c}{---}&\multicolumn{2}{c}{---}&\multicolumn{2}{c}{---}&1.43&1.22&1.46&1.14&1.46&1.08&1.46&1.02\\
155&0.75&0.10&\multicolumn{2}{c}{---}&\multicolumn{2}{c}{---}&1.51&1.23&1.53&1.15&1.54&1.09&1.54&1.04&1.53&0.99\\
\hline
155&0.85&0.06&\multicolumn{2}{c}{---}&\multicolumn{2}{c}{---}&\multicolumn{2}{c}{---}&\multicolumn{2}{c}{---}&\multicolumn{2}{c}{---}&\multicolumn{2}{c}{---}&1.11&1.03\\
155&0.85&0.07&\multicolumn{2}{c}{---}&\multicolumn{2}{c}{---}&\multicolumn{2}{c}{---}&\multicolumn{2}{c}{---}&1.17&1.10&1.20&1.02&1.20&0.96\\
155&0.85&0.08&\multicolumn{2}{c}{---}&\multicolumn{2}{c}{---}&\multicolumn{2}{c}{---}&1.27&1.09&1.28&1.02&1.28&0.97&1.27&0.92\\
155&0.85&0.09&\multicolumn{2}{c}{---}&1.29&1.21&1.35&1.10&1.36&1.03&1.36&0.98&1.35&0.93&1.34&0.89\\
155&0.85&0.10&1.37&1.22&1.41&1.12&1.43&1.05&1.43&0.99&1.43&0.95&1.42&0.90&1.40&0.86\\
\hline
\hline
168&0.65&0.07&\multicolumn{2}{c}{---}&\multicolumn{2}{c}{---}&\multicolumn{2}{c}{---}&\multicolumn{2}{c}{---}&\multicolumn{2}{c}{---}&\multicolumn{2}{c}{---}&1.29&1.22\\
168&0.65&0.08&\multicolumn{2}{c}{---}&\multicolumn{2}{c}{---}&\multicolumn{2}{c}{---}&\multicolumn{2}{c}{---}&\multicolumn{2}{c}{---}&1.40&1.20&1.42&1.13\\
168&0.65&0.09&\multicolumn{2}{c}{---}&\multicolumn{2}{c}{---}&\multicolumn{2}{c}{---}&1.43&1.32&1.49&1.21&1.51&1.14&1.51&1.08\\
168&0.65&0.10&\multicolumn{2}{c}{---}&\multicolumn{2}{c}{---}&1.52&1.33&1.57&1.22&1.59&1.15&1.59&1.09&1.59&1.04\\
\hline
168&0.75&0.07&\multicolumn{2}{c}{---}&\multicolumn{2}{c}{---}&\multicolumn{2}{c}{---}&\multicolumn{2}{c}{---}&\multicolumn{2}{c}{---}&1.22&1.06&1.23&0.99\\
168&0.75&0.08&\multicolumn{2}{c}{---}&\multicolumn{2}{c}{---}&\multicolumn{2}{c}{---}&1.28&1.15&1.31&1.06&1.31&1.00&1.30&0.95\\
168&0.75&0.09&\multicolumn{2}{c}{---}&\multicolumn{2}{c}{---}&1.36&1.15&1.38&1.07&1.39&1.01&1.38&0.96&1.37&0.92\\
168&0.75&0.10&\multicolumn{2}{c}{---}&1.43&1.17&1.46&1.09&1.46&1.03&1.46&0.98&1.45&0.93&1.44&0.89\\
\hline
168&0.85&0.06&\multicolumn{2}{c}{---}&\multicolumn{2}{c}{---}&\multicolumn{2}{c}{---}&\multicolumn{2}{c}{---}&\multicolumn{2}{c}{---}&1.06&0.97&1.06&0.90\\
168&0.85&0.07&\multicolumn{2}{c}{---}&\multicolumn{2}{c}{---}&\multicolumn{2}{c}{---}&1.12&1.04&1.14&0.96&1.14&0.90&1.13&0.86\\
168&0.85&0.08&\multicolumn{2}{c}{---}&\multicolumn{2}{c}{---}&1.21&1.03&1.22&0.96&1.22&0.91&1.21&0.87&1.19&0.83\\
168&0.85&0.09&1.22&1.15&1.28&1.04&1.29&0.98&1.29&0.92&1.28&0.88&1.27&0.84&1.25&0.80\\
168&0.85&0.10&1.34&1.06&1.36&0.99&1.36&0.94&1.36&0.89&1.35&0.85&1.33&0.81&1.31&0.78\\
\hline
\hline
181&0.65&0.07&\multicolumn{2}{c}{---}&\multicolumn{2}{c}{---}&\multicolumn{2}{c}{---}&\multicolumn{2}{c}{---}&\multicolumn{2}{c}{---}&1.23&1.16&1.26&1.06\\
181&0.65&0.08&\multicolumn{2}{c}{---}&\multicolumn{2}{c}{---}&\multicolumn{2}{c}{---}&\multicolumn{2}{c}{---}&1.34&1.14&1.35&1.07&1.35&1.01\\
181&0.65&0.09&\multicolumn{2}{c}{---}&\multicolumn{2}{c}{---}&1.36&1.27&1.42&1.15&1.43&1.08&1.44&1.02&1.43&0.97\\
181&0.65&0.10&\multicolumn{2}{c}{---}&1.44&1.27&1.49&1.17&1.51&1.09&1.52&1.03&1.51&0.98&1.50&0.94\\
\hline
181&0.75&0.06&\multicolumn{2}{c}{---}&\multicolumn{2}{c}{---}&\multicolumn{2}{c}{---}&\multicolumn{2}{c}{---}&\multicolumn{2}{c}{---}&\multicolumn{2}{c}{---}&1.09&0.94\\
181&0.75&0.07&\multicolumn{2}{c}{---}&\multicolumn{2}{c}{---}&\multicolumn{2}{c}{---}&\multicolumn{2}{c}{---}&1.16&1.00&1.17&0.94&1.16&0.89\\
181&0.75&0.08&\multicolumn{2}{c}{---}&\multicolumn{2}{c}{---}&1.22&1.09&1.24&1.01&1.25&0.95&1.24&0.90&1.23&0.86\\
181&0.75&0.09&\multicolumn{2}{c}{---}&1.29&1.10&1.32&1.02&1.32&0.96&1.32&0.91&1.31&0.87&1.29&0.83\\
181&0.75&0.10&1.36&1.12&1.39&1.04&1.40&0.98&1.39&0.93&1.39&0.88&1.37&0.84&1.36&0.80\\
\hline
181&0.85&0.06&\multicolumn{2}{c}{---}&\multicolumn{2}{c}{---}&\multicolumn{2}{c}{---}&\multicolumn{2}{c}{---}&1.01&0.92&1.01&0.85&1.00&0.81\\
181&0.85&0.07&\multicolumn{2}{c}{---}&\multicolumn{2}{c}{---}&1.06&0.99&1.09&0.91&1.09&0.86&1.08&0.81&1.06&0.77\\
181&0.85&0.08&\multicolumn{2}{c}{---}&1.15&0.99&1.16&0.92&1.16&0.87&1.15&0.82&1.14&0.78&1.12&0.75\\
181&0.85&0.09&1.22&1.00&1.23&0.93&1.23&0.88&1.22&0.83&1.21&0.79&1.19&0.76&1.17&0.73\\
181&0.85&0.10&1.30&0.95&1.30&0.90&1.29&0.85&1.28&0.81&1.27&0.77&1.25&0.74&1.23&0.71\\
\end{tabular}

\caption{The values for $\protect\imag \lambda_t$ corresponding to the
two values of $\delta$ in table~\protect\ref{tab:delta}.}
\label{tab:imlat}
\end{table}

\begin{table}[htb]
\squeezetable
\begin{tabular}[tb]{@{\hspace{3pt}}c@{\hspace{3pt}}c@{\hspace{3pt}}c@{\hspace{6pt}}r@{\hspace{3pt}}r@{\hspace{6pt}}r@{\hspace{3pt}}r@{\hspace{6pt}}r@{\hspace{3pt}}r@{\hspace{6pt}}r@{\hspace{3pt}}r@{\hspace{6pt}}r@{\hspace{3pt}}r@{\hspace{6pt}}r@{\hspace{3pt}}r@{\hspace{6pt}}r@{\hspace{3pt}}r}
\multicolumn{3}{c}{$V_{cb}$}&\multicolumn{2}{c}{0.037}&\multicolumn{2}{c}{0.038}&\multicolumn{2}{c}{0.039}&\multicolumn{2}{c}{0.040}&\multicolumn{2}{c}{0.041}&\multicolumn{2}{c}{0.042}&\multicolumn{2}{c}{0.043}\\
\hline
$m_t^*$&$B_K$&$\left|\frac{V_{ub}}{V_{cb}}\right|$&\multicolumn{14}{c}{$\bar\rho$}\\\hline\hline
155&0.65&0.08&\multicolumn{2}{c}{---}&\multicolumn{2}{c}{---}&\multicolumn{2}{c}{---}&\multicolumn{2}{c}{---}&\multicolumn{2}{c}{---}&\multicolumn{2}{c}{---}&0.026&-0.175\\
155&0.65&0.09&\multicolumn{2}{c}{---}&\multicolumn{2}{c}{---}&\multicolumn{2}{c}{---}&\multicolumn{2}{c}{---}&-0.055&-0.139&0.060&-0.232&0.120&-0.273\\
155&0.65&0.10&\multicolumn{2}{c}{---}&\multicolumn{2}{c}{---}&\multicolumn{2}{c}{---}&-0.015&-0.208&0.081&-0.279&0.141&-0.318&0.186&-0.344\\
\hline
155&0.75&0.07&\multicolumn{2}{c}{---}&\multicolumn{2}{c}{---}&\multicolumn{2}{c}{---}&\multicolumn{2}{c}{---}&\multicolumn{2}{c}{---}&\multicolumn{2}{c}{---}&0.035&-0.151\\
155&0.75&0.08&\multicolumn{2}{c}{---}&\multicolumn{2}{c}{---}&\multicolumn{2}{c}{---}&\multicolumn{2}{c}{---}&-0.002&-0.149&0.078&-0.212&0.127&-0.245\\
155&0.75&0.09&\multicolumn{2}{c}{---}&\multicolumn{2}{c}{---}&\multicolumn{2}{c}{---}&0.036&-0.211&0.106&-0.262&0.155&-0.292&0.192&-0.313\\
155&0.75&0.10&\multicolumn{2}{c}{---}&\multicolumn{2}{c}{---}&0.055&-0.258&0.125&-0.306&0.176&-0.336&0.215&-0.357&0.247&-0.373\\
\hline
155&0.85&0.06&\multicolumn{2}{c}{---}&\multicolumn{2}{c}{---}&\multicolumn{2}{c}{---}&\multicolumn{2}{c}{---}&\multicolumn{2}{c}{---}&\multicolumn{2}{c}{---}&0.006&-0.097\\
155&0.85&0.07&\multicolumn{2}{c}{---}&\multicolumn{2}{c}{---}&\multicolumn{2}{c}{---}&\multicolumn{2}{c}{---}&-0.014&-0.107&0.068&-0.175&0.113&-0.207\\
155&0.85&0.08&\multicolumn{2}{c}{---}&\multicolumn{2}{c}{---}&\multicolumn{2}{c}{---}&0.043&-0.184&0.105&-0.229&0.148&-0.257&0.180&-0.276\\
155&0.85&0.09&\multicolumn{2}{c}{---}&-0.041&-0.145&0.072&-0.236&0.132&-0.276&0.175&-0.303&0.209&-0.322&0.237&-0.336\\
155&0.85&0.10&-0.014&-0.202&0.088&-0.279&0.151&-0.319&0.197&-0.346&0.233&-0.365&0.263&-0.379&0.288&-0.391\\
\hline
\hline
168&0.65&0.07&\multicolumn{2}{c}{---}&\multicolumn{2}{c}{---}&\multicolumn{2}{c}{---}&\multicolumn{2}{c}{---}&\multicolumn{2}{c}{---}&\multicolumn{2}{c}{---}&-0.028&-0.099\\
168&0.65&0.08&\multicolumn{2}{c}{---}&\multicolumn{2}{c}{---}&\multicolumn{2}{c}{---}&\multicolumn{2}{c}{---}&\multicolumn{2}{c}{---}&0.036&-0.184&0.098&-0.229\\
168&0.65&0.09&\multicolumn{2}{c}{---}&\multicolumn{2}{c}{---}&\multicolumn{2}{c}{---}&-0.039&-0.155&0.067&-0.239&0.125&-0.278&0.168&-0.303\\
168&0.65&0.10&\multicolumn{2}{c}{---}&\multicolumn{2}{c}{---}&-0.011&-0.214&0.084&-0.284&0.145&-0.322&0.190&-0.347&0.226&-0.366\\
\hline
168&0.75&0.07&\multicolumn{2}{c}{---}&\multicolumn{2}{c}{---}&\multicolumn{2}{c}{---}&\multicolumn{2}{c}{---}&\multicolumn{2}{c}{---}&0.044&-0.159&0.097&-0.198\\
168&0.75&0.08&\multicolumn{2}{c}{---}&\multicolumn{2}{c}{---}&\multicolumn{2}{c}{---}&0.005&-0.157&0.083&-0.217&0.131&-0.249&0.167&-0.271\\
168&0.75&0.09&\multicolumn{2}{c}{---}&\multicolumn{2}{c}{---}&0.038&-0.215&0.109&-0.265&0.158&-0.295&0.195&-0.316&0.225&-0.332\\
168&0.75&0.10&\multicolumn{2}{c}{---}&0.054&-0.260&0.126&-0.308&0.177&-0.338&0.217&-0.360&0.250&-0.375&0.277&-0.388\\
\hline
168&0.85&0.06&\multicolumn{2}{c}{---}&\multicolumn{2}{c}{---}&\multicolumn{2}{c}{---}&\multicolumn{2}{c}{---}&\multicolumn{2}{c}{---}&0.018&-0.108&0.075&-0.154\\
168&0.85&0.07&\multicolumn{2}{c}{---}&\multicolumn{2}{c}{---}&\multicolumn{2}{c}{---}&-0.005&-0.116&0.073&-0.179&0.118&-0.211&0.150&-0.231\\
168&0.85&0.08&\multicolumn{2}{c}{---}&\multicolumn{2}{c}{---}&0.045&-0.187&0.108&-0.232&0.151&-0.260&0.183&-0.279&0.209&-0.293\\
168&0.85&0.09&-0.049&-0.141&0.071&-0.237&0.132&-0.278&0.177&-0.305&0.211&-0.324&0.239&-0.338&0.262&-0.349\\
168&0.85&0.10&0.085&-0.279&0.149&-0.320&0.197&-0.347&0.234&-0.367&0.265&-0.381&0.290&-0.392&0.311&-0.401\\
\hline
\hline
181&0.65&0.07&\multicolumn{2}{c}{---}&\multicolumn{2}{c}{---}&\multicolumn{2}{c}{---}&\multicolumn{2}{c}{---}&\multicolumn{2}{c}{---}&-0.019&-0.109&0.064&-0.177\\
181&0.65&0.08&\multicolumn{2}{c}{---}&\multicolumn{2}{c}{---}&\multicolumn{2}{c}{---}&\multicolumn{2}{c}{---}&0.038&-0.187&0.100&-0.231&0.143&-0.259\\
181&0.65&0.09&\multicolumn{2}{c}{---}&\multicolumn{2}{c}{---}&-0.047&-0.150&0.065&-0.239&0.126&-0.279&0.169&-0.305&0.204&-0.323\\
181&0.65&0.10&\multicolumn{2}{c}{---}&-0.021&-0.208&0.081&-0.283&0.143&-0.322&0.190&-0.348&0.227&-0.367&0.257&-0.381\\
\hline
181&0.75&0.06&\multicolumn{2}{c}{---}&\multicolumn{2}{c}{---}&\multicolumn{2}{c}{---}&\multicolumn{2}{c}{---}&\multicolumn{2}{c}{---}&\multicolumn{2}{c}{---}&0.053&-0.139\\
181&0.75&0.07&\multicolumn{2}{c}{---}&\multicolumn{2}{c}{---}&\multicolumn{2}{c}{---}&\multicolumn{2}{c}{---}&0.045&-0.161&0.099&-0.200&0.136&-0.224\\
181&0.75&0.08&\multicolumn{2}{c}{---}&\multicolumn{2}{c}{---}&0.000&-0.155&0.082&-0.217&0.131&-0.250&0.168&-0.272&0.197&-0.288\\
181&0.75&0.09&\multicolumn{2}{c}{---}&0.031&-0.212&0.106&-0.264&0.156&-0.296&0.195&-0.317&0.225&-0.333&0.250&-0.345\\
181&0.75&0.10&0.043&-0.255&0.120&-0.306&0.174&-0.338&0.216&-0.360&0.249&-0.376&0.277&-0.388&0.300&-0.398\\
\hline
181&0.85&0.06&\multicolumn{2}{c}{---}&\multicolumn{2}{c}{---}&\multicolumn{2}{c}{---}&\multicolumn{2}{c}{---}&0.019&-0.110&0.077&-0.156&0.112&-0.181\\
181&0.85&0.07&\multicolumn{2}{c}{---}&\multicolumn{2}{c}{---}&-0.011&-0.113&0.072&-0.180&0.118&-0.212&0.151&-0.232&0.176&-0.248\\
181&0.85&0.08&\multicolumn{2}{c}{---}&0.038&-0.184&0.105&-0.232&0.149&-0.260&0.183&-0.279&0.210&-0.294&0.231&-0.305\\
181&0.85&0.09&0.062&-0.233&0.128&-0.277&0.174&-0.304&0.210&-0.324&0.239&-0.338&0.262&-0.349&0.282&-0.358\\
181&0.85&0.10&0.142&-0.318&0.193&-0.346&0.232&-0.366&0.263&-0.381&0.289&-0.393&0.311&-0.402&0.329&-0.409\\
\end{tabular}

\caption{The values for $\bar\rho$ corresponding to the
two values of $\delta$ in table~\protect\ref{tab:delta}.}
\label{tab:rhobar}
\end{table}

\begin{table}[htb]
\squeezetable
\begin{tabular}[tb]{@{\hspace{3pt}}c@{\hspace{3pt}}c@{\hspace{3pt}}c@{\hspace{6pt}}r@{\hspace{3pt}}r@{\hspace{6pt}}r@{\hspace{3pt}}r@{\hspace{6pt}}r@{\hspace{3pt}}r@{\hspace{6pt}}r@{\hspace{3pt}}r@{\hspace{6pt}}r@{\hspace{3pt}}r@{\hspace{6pt}}r@{\hspace{3pt}}r@{\hspace{6pt}}r@{\hspace{3pt}}r}
\multicolumn{3}{c}{$V_{cb}$}&\multicolumn{2}{c}{0.037}&\multicolumn{2}{c}{0.038}&\multicolumn{2}{c}{0.039}&\multicolumn{2}{c}{0.040}&\multicolumn{2}{c}{0.041}&\multicolumn{2}{c}{0.042}&\multicolumn{2}{c}{0.043}\\
\hline
$m_t^*$&$B_K$&$\left|\frac{V_{ub}}{V_{cb}}\right|$&\multicolumn{14}{c}{$\bar\eta$}\\\hline\hline
155&0.65&0.08&\multicolumn{2}{c}{---}&\multicolumn{2}{c}{---}&\multicolumn{2}{c}{---}&\multicolumn{2}{c}{---}&\multicolumn{2}{c}{---}&\multicolumn{2}{c}{---}&0.352&0.307\\
155&0.65&0.09&\multicolumn{2}{c}{---}&\multicolumn{2}{c}{---}&\multicolumn{2}{c}{---}&\multicolumn{2}{c}{---}&0.394&0.373&0.393&0.323&0.379&0.290\\
155&0.65&0.10&\multicolumn{2}{c}{---}&\multicolumn{2}{c}{---}&\multicolumn{2}{c}{---}&0.441&0.390&0.434&0.342&0.419&0.307&0.400&0.278\\
\hline
155&0.75&0.07&\multicolumn{2}{c}{---}&\multicolumn{2}{c}{---}&\multicolumn{2}{c}{---}&\multicolumn{2}{c}{---}&\multicolumn{2}{c}{---}&\multicolumn{2}{c}{---}&0.307&0.270\\
155&0.75&0.08&\multicolumn{2}{c}{---}&\multicolumn{2}{c}{---}&\multicolumn{2}{c}{---}&\multicolumn{2}{c}{---}&0.353&0.320&0.345&0.283&0.330&0.255\\
155&0.75&0.09&\multicolumn{2}{c}{---}&\multicolumn{2}{c}{---}&\multicolumn{2}{c}{---}&0.396&0.337&0.383&0.299&0.366&0.270&0.348&0.245\\
155&0.75&0.10&\multicolumn{2}{c}{---}&\multicolumn{2}{c}{---}&0.438&0.358&0.423&0.319&0.405&0.287&0.385&0.260&0.366&0.237\\
\hline
155&0.85&0.06&\multicolumn{2}{c}{---}&\multicolumn{2}{c}{---}&\multicolumn{2}{c}{---}&\multicolumn{2}{c}{---}&\multicolumn{2}{c}{---}&\multicolumn{2}{c}{---}&0.265&0.247\\
155&0.85&0.07&\multicolumn{2}{c}{---}&\multicolumn{2}{c}{---}&\multicolumn{2}{c}{---}&\multicolumn{2}{c}{---}&0.309&0.290&0.301&0.255&0.288&0.230\\
155&0.85&0.08&\multicolumn{2}{c}{---}&\multicolumn{2}{c}{---}&\multicolumn{2}{c}{---}&0.351&0.302&0.337&0.269&0.321&0.243&0.304&0.221\\
155&0.85&0.09&\multicolumn{2}{c}{---}&0.395&0.370&0.391&0.320&0.375&0.286&0.357&0.258&0.338&0.234&0.319&0.213\\
155&0.85&0.10&0.441&0.393&0.433&0.342&0.415&0.305&0.395&0.275&0.375&0.249&0.354&0.227&0.335&0.207\\
\hline
\hline
168&0.65&0.07&\multicolumn{2}{c}{---}&\multicolumn{2}{c}{---}&\multicolumn{2}{c}{---}&\multicolumn{2}{c}{---}&\multicolumn{2}{c}{---}&\multicolumn{2}{c}{---}&0.308&0.293\\
168&0.65&0.08&\multicolumn{2}{c}{---}&\multicolumn{2}{c}{---}&\multicolumn{2}{c}{---}&\multicolumn{2}{c}{---}&\multicolumn{2}{c}{---}&0.351&0.301&0.339&0.269\\
168&0.65&0.09&\multicolumn{2}{c}{---}&\multicolumn{2}{c}{---}&\multicolumn{2}{c}{---}&0.396&0.366&0.392&0.318&0.377&0.285&0.360&0.258\\
168&0.65&0.10&\multicolumn{2}{c}{---}&\multicolumn{2}{c}{---}&0.442&0.386&0.433&0.339&0.417&0.303&0.399&0.274&0.379&0.249\\
\hline
168&0.75&0.07&\multicolumn{2}{c}{---}&\multicolumn{2}{c}{---}&\multicolumn{2}{c}{---}&\multicolumn{2}{c}{---}&\multicolumn{2}{c}{---}&0.306&0.265&0.293&0.238\\
168&0.75&0.08&\multicolumn{2}{c}{---}&\multicolumn{2}{c}{---}&\multicolumn{2}{c}{---}&0.353&0.317&0.343&0.279&0.328&0.251&0.311&0.227\\
168&0.75&0.09&\multicolumn{2}{c}{---}&\multicolumn{2}{c}{---}&0.396&0.334&0.382&0.296&0.365&0.266&0.346&0.241&0.328&0.219\\
168&0.75&0.10&\multicolumn{2}{c}{---}&0.438&0.357&0.423&0.317&0.404&0.284&0.384&0.257&0.364&0.233&0.344&0.213\\
\hline
168&0.85&0.06&\multicolumn{2}{c}{---}&\multicolumn{2}{c}{---}&\multicolumn{2}{c}{---}&\multicolumn{2}{c}{---}&\multicolumn{2}{c}{---}&0.264&0.242&0.254&0.216\\
168&0.85&0.07&\multicolumn{2}{c}{---}&\multicolumn{2}{c}{---}&\multicolumn{2}{c}{---}&0.309&0.287&0.300&0.252&0.286&0.227&0.270&0.205\\
168&0.85&0.08&\multicolumn{2}{c}{---}&\multicolumn{2}{c}{---}&0.350&0.300&0.336&0.266&0.320&0.240&0.302&0.217&0.285&0.198\\
168&0.85&0.09&0.394&0.372&0.391&0.319&0.375&0.284&0.356&0.255&0.337&0.231&0.317&0.210&0.299&0.192\\
168&0.85&0.10&0.433&0.343&0.416&0.304&0.395&0.273&0.374&0.247&0.353&0.224&0.333&0.204&0.314&0.186\\
\hline
\hline
181&0.65&0.07&\multicolumn{2}{c}{---}&\multicolumn{2}{c}{---}&\multicolumn{2}{c}{---}&\multicolumn{2}{c}{---}&\multicolumn{2}{c}{---}&0.309&0.290&0.302&0.254\\
181&0.65&0.08&\multicolumn{2}{c}{---}&\multicolumn{2}{c}{---}&\multicolumn{2}{c}{---}&\multicolumn{2}{c}{---}&0.351&0.300&0.339&0.267&0.323&0.241\\
181&0.65&0.09&\multicolumn{2}{c}{---}&\multicolumn{2}{c}{---}&0.395&0.368&0.392&0.318&0.377&0.283&0.359&0.256&0.341&0.232\\
181&0.65&0.10&\multicolumn{2}{c}{---}&0.441&0.390&0.434&0.339&0.418&0.302&0.399&0.272&0.379&0.247&0.359&0.224\\
\hline
181&0.75&0.06&\multicolumn{2}{c}{---}&\multicolumn{2}{c}{---}&\multicolumn{2}{c}{---}&\multicolumn{2}{c}{---}&\multicolumn{2}{c}{---}&\multicolumn{2}{c}{---}&0.260&0.225\\
181&0.75&0.07&\multicolumn{2}{c}{---}&\multicolumn{2}{c}{---}&\multicolumn{2}{c}{---}&\multicolumn{2}{c}{---}&0.306&0.264&0.293&0.236&0.278&0.213\\
181&0.75&0.08&\multicolumn{2}{c}{---}&\multicolumn{2}{c}{---}&0.353&0.318&0.344&0.279&0.328&0.250&0.311&0.226&0.293&0.205\\
181&0.75&0.09&\multicolumn{2}{c}{---}&0.396&0.337&0.383&0.297&0.365&0.266&0.346&0.240&0.327&0.218&0.308&0.198\\
181&0.75&0.10&0.439&0.361&0.425&0.318&0.406&0.285&0.385&0.256&0.364&0.232&0.344&0.211&0.324&0.193\\
\hline
181&0.85&0.06&\multicolumn{2}{c}{---}&\multicolumn{2}{c}{---}&\multicolumn{2}{c}{---}&\multicolumn{2}{c}{---}&0.264&0.241&0.254&0.214&0.240&0.193\\
181&0.85&0.07&\multicolumn{2}{c}{---}&\multicolumn{2}{c}{---}&0.309&0.288&0.301&0.252&0.286&0.226&0.270&0.204&0.254&0.185\\
181&0.85&0.08&\multicolumn{2}{c}{---}&0.351&0.302&0.337&0.267&0.320&0.240&0.302&0.216&0.284&0.197&0.267&0.179\\
181&0.85&0.09&0.393&0.322&0.376&0.285&0.357&0.256&0.337&0.231&0.318&0.209&0.299&0.190&0.280&0.174\\
181&0.85&0.10&0.418&0.307&0.397&0.274&0.376&0.247&0.354&0.223&0.334&0.203&0.314&0.185&0.294&0.169\\
\end{tabular}

\caption{The values for $\bar\eta$ corresponding to the
two values of $\delta$ in table~\protect\ref{tab:delta}.}
\label{tab:etabar}
\end{table}

\section{A 1995 look at the \protect${\text{\bf K}_{\text{\bf L}}
         \!-\!\text{\bf K}_{\text{\bf S}}}\,$-mass difference\/}\label{secmass}
In this section we will have a look at the status of the \kkmd\/ $\Delta
m_K$.

The short distance part of $\Delta m_K$, denoted by $(\Delta
m_K)_{\text{SD}}$, reads
\begin{eqnarray}
\frac{\left(\Delta m_K\right)_{\text{SD}}}{m_K} &=&
\frac{G_F^2}{6\pi^2} f_K^2 B_K M_W^2 \\ \nonumber
 & & \cdot
\left[
	\left(\real \lambda_c\right)^2 x_c^{\star} \eta_1^{\star}
	+2 \left(\real \lambda_c\right) \left(\real \lambda_t\right)
	S\left(x_c^{\star},x_t^{\star}\right) \eta_3^{\star}
	+\left(\real \lambda_t\right)^2 S\left(x_t^{\star}\right) \eta_2^{\star}
\right] ,
\end{eqnarray}
where the small imaginary parts of $\lambda_c$ and $\lambda_t$ have
been neglected.  The three terms in the brackets contribute roughly in
the ratio 100:10:1, therefore the term containing $\eta_1^{\star}$ is
most important, the one with $\eta_2^{\star}$ is least.

Because $\eta_1^{\star}$ strongly depends on its input parameters,
especially on $m_c^{\star}$ and $\laMSb$, it does not make sense to
use the constant defined in (\ref{etas}).  We therefore calculate
$\eta_1^{\star}$ for each set of parameters in our numerical evaluation.
Inserting our standard set of values defined in sect.~\ref{par}, we
obtain
\begin{eqnarray}
\frac{\left(\Delta m_K\right)_{\text{SD}}}
	{\left(\Delta m_K\right)_{\text{exp}}}
&=&
\left\{
\renewcommand{\arraystretch}{1.5}
\begin{array}{rl@{\hspace{1cm}}rl}
0.52 & {+0.17 \atop -0.11} & \text{for} &
\laMSb=0.210\gev \\
0.67 & {+0.25 \atop -0.14} & \text{for} &
\laMSb=0.310\gev \\
0.91 & {+0.39 \atop -0.20} & \text{for} &
\laMSb=0.410\gev
\end{array}
\right.
\label{dmkresult}
\end{eqnarray}
The errors are estimated by a scan through the allowed parameter space
and includes the error stemming from scale variations in the
$\eta_i^{\star}$'s.

\begin{figure}[htb]
\centerline{
\epsffile{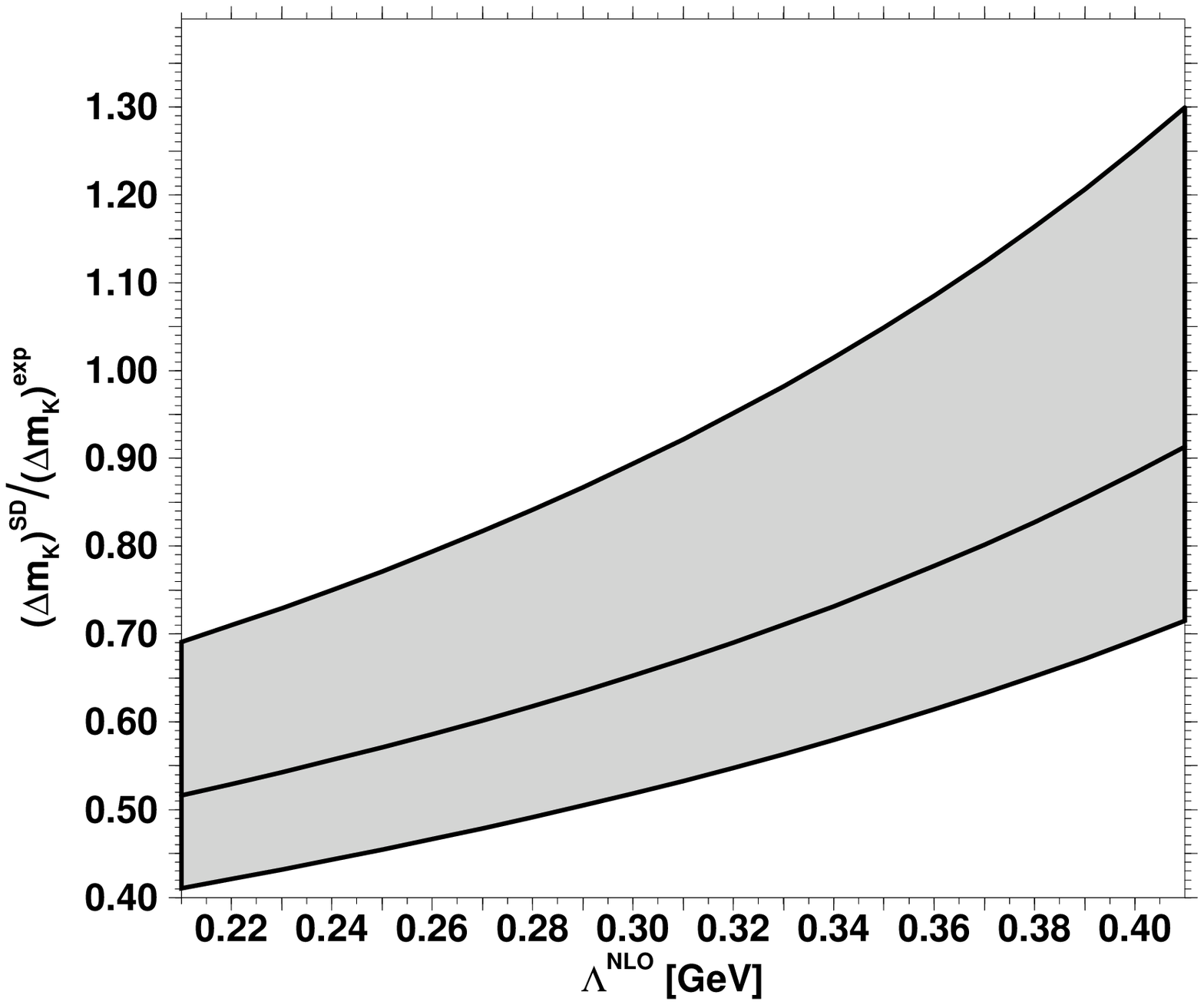}
}
\caption{The dependence of the ratio of the short distance part of the
\kkm\/ to the experimentally measured value on $\Lambda^{\text{NLO}}
\equiv \protect\laMSb$.  The curve in the middle is obtained by
choosing the central values of the parameters as given in
sect.~\protect\ref{par}.  All values lying in the shaded area are
compatible within the error bands quoted in this section.  The
increasing height of the band displaying the error is due to the
growing scale uncertainties present in the coefficient
$\eta_1^{\star}$.}
\label{fig:dmk-la}
\end{figure}

The strong $\laMSb$ dependence of $\left(\Delta
m_K\right)_{\text{SD}}/\left(\Delta m_K\right)_{\text{exp}}$ has been
visualized in fig.~\ref{fig:dmk-la}.  The central line is obtained by
using the central values defined in sect.~\ref{par}, the band shaded
in grey displays the error.

For large values of $\laMSb$ the uncertainties in $\eta^{\star}_1$ due to
scale variations become large indicating the breakdown of
perturbation theory.  Therefore the error bar on $\Delta m_K$ which is
then dominated by this scale uncertainty grows very large prohibiting
a precise prediction for the mass difference.  One will have to see,
whether in the future $\laMSb$ will continue growing in the future and
thereby bringing the next-to-leading order result for $\eta^{\star}_1$ into
troubles.

Let us now discuss the differences between our new result and previous
analyses: In most textbooks $\Delta m_K$ is termed to be dominated by
poorly calculable long-distance physics.  Yet by power counting
arguments, long-distance effects should be suppressed by a power of
$\Lambda_{\text{QCD}} ^2/m_c^{\star \, 2}$ with respect to the short
distance part because the coefficient of the leading dimension six
operator contributing to the $\eta^{\star}_1$-part of the effective
Hamiltonian in (\ref{s2}) is proportional to $m_c^{\star \, 2}$ (see
e.g.\ \cite{Shif}).

A short look at (\ref{dmkresult}) clearly exhibits a short distance
dominance. Let us discuss the steps which have guided us to this
result.

Already our 1993 analysis \cite{hn1}, in which we have calculated the
coefficient $\eta^{\star}_1$ in the next-to-leading order approximation,
has resulted in a large enhancement of the theoretical prediction for the
\kkm.  This fact is true, because
\begin{itemize}
\item
the next-to-leading order correction have largely increased the value
of $\eta^{\star}_1$ and
\item
the experimental value  for $\laMSb$ has risen in the last
decade.
\end{itemize}
Both findings lead to the drastic increase of $\eta_1^{\star}$ by
approximately 65\%, which we get by comparing (\ref{old}) and
(\ref{etas}).

Finally, our new analysis compared to \cite{hn1} for the first time
uses the coefficient $\eta_3^{\star}$ calculated in the
next-to-leading approximation.  This quantity again enlarges the
result for $\Delta m_K$.  Because $\laMSb$ has grown again in the
meantime thereby enlarging the theoretical prediction once more, we are
now able to reproduce the experimentally measured value to 50--100\%
by short distance physics.

Some authors attributed the deficit in $\Delta m_K$ to new physics.
The large scale uncertainties present in the coefficient
$\eta_1^{\star}$, which obscure a clean determination of the Standard Model
contribution, make the \kkm\/ a poor laboratory to search for the
impact of new physics.

\section*{Acknowledgements}
We thank Andrzej Buras for suggesting the topic and permanent
encouragement. We have enjoyed many useful discussions with him and
Gaby Ostermaier.  U.N.\ thanks Patricia Ball for her explanations of
the determination of $V_{cb}$ in \cite{bbb} and Ikarus Bigi for
discussions on the same topic.  S.H.\ thanks Fred Jegerlehner for
interesting discussions.


\begin{references}
\bibitem{ccft}  J.~H.~Christenson, J.~W.~Cronin,
             V.~L.~Fitch and R.~Turlay,
             Phys.~Rev.~Lett.~13 (1964) 138.\\
             J.~H.~Christenson, J.~W.~Cronin,
             V.~L.~Fitch and R.~Turlay,
             Phys.~Rev.~140B (1965) 74.
\bibitem{cbh} L.~Chau, Phys.~Rep.~95 (1983) 1. \
              A.~J.~Buras and M.~K.~Harlander,
               {\sl A Top-Quark Story: Quark Mixing, CP-Violation and
               Rare Decays in the Standard Model}, in
               {\sl Heavy Flavours}, ed.~A.~J.~Buras and M.~Lindner,
               World Scientific, Singapore, 1993.
\bibitem{il} T.~Inami and C.~S.~Lim,
             Progr.~Theor.~Phys.~65 (1981)  297
             [Erratum: 65 (1981) 1772].
\bibitem{gw} F.~J.~Gilman and M.~B.~Wise,
             Phys.~Rev.~D27 (1983) 1128.
\bibitem{fp} J.~M.~Flynn, Mod.~Phys.~Lett.~A5 (1990) 877.\\
             A.~Datta, J.~Fr\"ohlich and E.~A.~Paschos,
             Z.~Phys.~C46 (1990) 63.
\bibitem{hn1}  S.~Herrlich and U.~Nierste,
              \np B419 (1994) 292.
\bibitem{bjw} A.~J.~Buras, M.~Jamin and
              P.~H.~Weisz, Nucl.~Phys.~B347 (1990) 491.
\bibitem{hn86}  S.~Herrlich and U.~Nierste,
             {\sl The complete $|\Delta \text{S}|\text{=2}$--hamiltonian in
               next-to-lead\-ing order},
            \prp\/ TUM-T31-86/95 in preparation.
\bibitem{pdg} Particle Data Group,
              \pr D50 (1994) 1173.
\bibitem{blo} A.~J.~Buras, M.~E.~Lautenbacher and G.~Ostermaier, \pr D50
              (1994) 3433.
\bibitem{w} L.~Wolfenstein, \prl\/ 51 (1983) 1945.
\bibitem{atom} A.~J.~Buras, M.~Jamin and
              P.~H.~Weisz, Nucl.~Phys.~B408 (1993) 209.
\bibitem{bbb} P.~Ball, M.~Beneke and V.M.~Braun, preprint
              CERN-TH/95-65,hep-ph/9503492.
\bibitem{bigi} I.~Bigi, Plenary talk at the DPG conference, Karlsruhe 1995.
\bibitem{bn}  P.~Ball and U.~Nierste, \pr\/
              D50 (1994) 5841.
\bibitem{p}  O.~Podobrin, Plenary talk at the DPG conference, Karlsruhe 1995.
\bibitem{cdf} F.~Abe et al., CDF, \pr D50 (1994) 2966, \prl 73 (1994) 225.
\bibitem{srie} S.~Riemann, Plenary talk at the DPG conference, Karlsruhe 1995.
\bibitem{d0} S.~Abachi et al., D0, FERMILAB-PUB 1995.
\bibitem{sop} D.~E.~Soper, {\sl Summary of the XXX Recontre de
              Moriond, QCD Session}, hep-ph/9506218.
\bibitem{bbg} W.~A.~Bardeen, A.~J.~Buras and J.-M.~G\'erard,
    \pl\/  B211 (1988) 343. \\
    J.-M.~G\'erard, Acta Phys.~Pol.~B21 (1990) 257.
\bibitem{gdh} J.~F.~Donoghue, E.~Golowich and B.~R.~Holstein,
       \pl\/ B119 (1982) 412.
\bibitem{pr} A.~Pich and  E.~de Rafael, \pl\/ B158 (1985) 477.\\
         J.~Prades et al., Z.~Phys.~C51 (1991) 287.
\bibitem{bp} J.~Bijnens and J.~Prades, NORDITA-95/11,
      hep-ph/9502363.
\bibitem{latt} M.~Crisafulli et al., hep-lat/9505020.
\bibitem{i} N.~Ishizuka et al., Phys.~Rev.~Lett.~71 (1993) 24.
\bibitem{beth}  S.~Bethke, Proc.~of the Summer School on Hadronic
           Aspects of Collider Physics, Zuoz, Switzerland, August 1994.
\bibitem{def} A.~Duncan, E.~Eichten, J.~Flynn, B.~Hill, G.~Hockney and
            H.~Thacker, preprint FERMILAB-PUB-94/164-T,
            hep-lat/9407025. \\
            C.W.~Bernard, J.N.~Labrenz and A.~Soni, \prd\/ D49 (1994)
            2536.\\
            T.~Draper and C.~McNeile, \np\/ (Proc.~Suppl.) 34 (1994) 453.
\bibitem{bpbd} E.~Bagan, P.~Ball, V.M.~Braun and H.G.~Dosch, \pl\/
              B278 (1992) 457. \\
             M.~Neubert, \pr\/ D45 (1992) 2451.
\bibitem{ajb} A.~J.~Buras, \pl B317 (1993) 449.
\bibitem{ali} A.~Ali and D.~London, preprint DESY-93-022,
               to be publ. in {\sl Proc. of ECFA Workshop on the
               Physics of a B Meson Factory, ed. R.~Aleksan, A.~Ali
               (1993)}.
\bibitem{jak} K.~Jakobs, ALEPH, Talk at the DPG conference, Karlsruhe 1995.
\bibitem{Shif}  M.~A.~Shifman, Int.\ J.\ of Mod.\ Phys.\ A12 (1988)
                2769.
\end{references}
\end{document}